\documentclass[12pt,twocolumn]{openjournal}
\usepackage{xcolor}
\usepackage{textgreek}
\usepackage[utf8]{inputenc}
\usepackage[english]{babel}
\newcommand{\xmm}{\textit{XMM-Newton}\xspace}
\usepackage{hyperref}
\usepackage{rotating}
\hypersetup{
    unicode, 
    colorlinks=true,
    linkcolor=linkcolor,
    citecolor=linkcolor,
    filecolor=linkcolor,
    urlcolor=linkcolor,
}
\usepackage{color,colortbl}
\definecolor{linkcolor}{rgb}{0.0,0.3,0.5}
\usepackage{tensind}
\tensordelimiter{?}
\DeclareGraphicsExtensions{.bmp,.png,.jpg,.pdf}
\usepackage{verbatim}
\usepackage[normalem]{ulem}
\usepackage{orcidlink}
\usepackage{soul}
\usepackage{amsmath}
\usepackage{xspace}
\usepackage{threeparttable}
\usepackage{tabularx}
\graphicspath{ {./Figures/} }
\usepackage{tikz}

\begin{document}
\title{SDSS-C4\,3028: the Nearest Blue Galaxy Cluster Devoid of an Intracluster Medium}
\author{
Shweta Jain\orcidlink{0009-0009-1792-7199}\altaffilmark{1,*},
Yuanyuan Su\orcidlink{0000-0002-3886-1258}\altaffilmark{1,*},
Andra Stroe\orcidlink{0000-0001-8322-4162}\altaffilmark{2,3},
Paul Nulsen\orcidlink{0000-0003-0297-4493}\altaffilmark{2,4},
Hyejeon Cho\orcidlink{0000-0001-5966-5072}\altaffilmark{5,6},
Kim HyeongHan\orcidlink{0000-0002-2550-5545}\altaffilmark{5},\\
M. James Jee\orcidlink{0000-0002-5751-3697}\altaffilmark{5,7},
Ralph P. Kraft\orcidlink{0000-0002-0765-0511}\altaffilmark{2},
Scott Randall\orcidlink{0000-0002-3984-4337}\altaffilmark{2},
Jimmy A. Irwin\orcidlink{0000-0003-4307-8521}\altaffilmark{8},
Ryan L. Sanders\orcidlink{0000-0003-4792-9119}\altaffilmark{1},\\
Christine Jones\orcidlink{0000-0003-2206-4243}\altaffilmark{2}
}
\altaffiltext{1}{Department of Physics and Astronomy, University of Kentucky, Lexington, KY 40506, USA}
\altaffiltext{2}{Center for Astrophysics $\vert$ Harvard \& Smithsonian, 60 Garden Street, Cambridge, MA 02138, USA}
\altaffiltext{3}{Space Telescope Science Institute, 3700 San Martin Drive, Baltimore, MD 21218, USA}
\altaffiltext{4}{ICRAR, University of Western Australia, 35 Stirling Hwy., Crawley, WA 6009, Australia}
\altaffiltext{5}{Department of Astronomy, Yonsei University, Seoul 03722, Republic of Korea}
\altaffiltext{6}{Center for Galaxy Evolution Research, Yonsei University, Seoul 03722, Republic of Korea}
\altaffiltext{7}{Department of Physics and Astronomy, University of California, Davis, CA 95616, USA}
\altaffiltext{8}{Department of Physics and Astronomy, University of Alabama, Tuscaloosa, AL 35487, USA}

\altaffiltext{*}{Email: \href{mailto:sja281@uky.edu}{\texttt{sja281@uky.edu}}, \href{mailto:ysu262@g.uky.edu}{\texttt{ysu262@g.uky.edu}}}

\begin{abstract}
    SDSS-C4\,3028 is a galaxy cluster at $z=0.061$, notable for its unusually high fraction of star-forming galaxies with 19 star-forming and 11 quiescent spectroscopically-confirmed member galaxies. From Subaru/HSC imaging, we derived a weak lensing mass of $M_{200} = (1.3 \pm 0.9) \times 10^{14}\, \rm M_\odot$, indicating a low-mass cluster. This is in excellent agreement with its dynamical mass of $M_{200} = (1.0\pm0.4)\times10^{14} \rm M_\odot$, derived from SDSS spectroscopic data. \xmm observations reveal that its X-ray emission is uniform and fully consistent with the astrophysical X-ray background, with no evidence for an intracluster medium (ICM). The 3$\sigma$ upper limit of $L_{\rm X}(0.1-2.4\rm\,keV)=7.7\times10^{42}$\,erg s$^{-1}$ on the cluster's X-ray luminosity falls below the value expected from the $L_{\rm X}-M_{\rm halo}$ scaling relation of nearby galaxy clusters.
    We derived star-formation histories for its member galaxies using the photometric spectral energy distribution from SDSS, 2MASS, and \textit{WISE} data. Most of its quiescent galaxies reside within the central 300\,kpc, while star-forming ones dominate the outer region (300\,kpc -- 1\,Mpc). The core region has formed the bulk of its stellar mass approximately 1.5\,Gyr earlier than the outskirts. We infer a long quenching time of $>3$\,Gyr for its quiescent galaxies, consistent with slow quenching mechanisms such as galaxy-galaxy interaction or strangulation. 
    These findings suggest that SDSS-C4\,3028 may have undergone an “inside-out” formation and quenching process. Its ICM may have been expelled by intense AGN feedback after core formation but before full cluster assembly. The high fraction ($\sim$0.63) of star-forming members likely results from the absence of ram pressure stripping in this blue cluster, supporting the important role of ram pressure stripping in quenching galaxies in clusters. 
\end{abstract}


\maketitle

\begin{figure}[ht!]
\includegraphics[width=\linewidth]{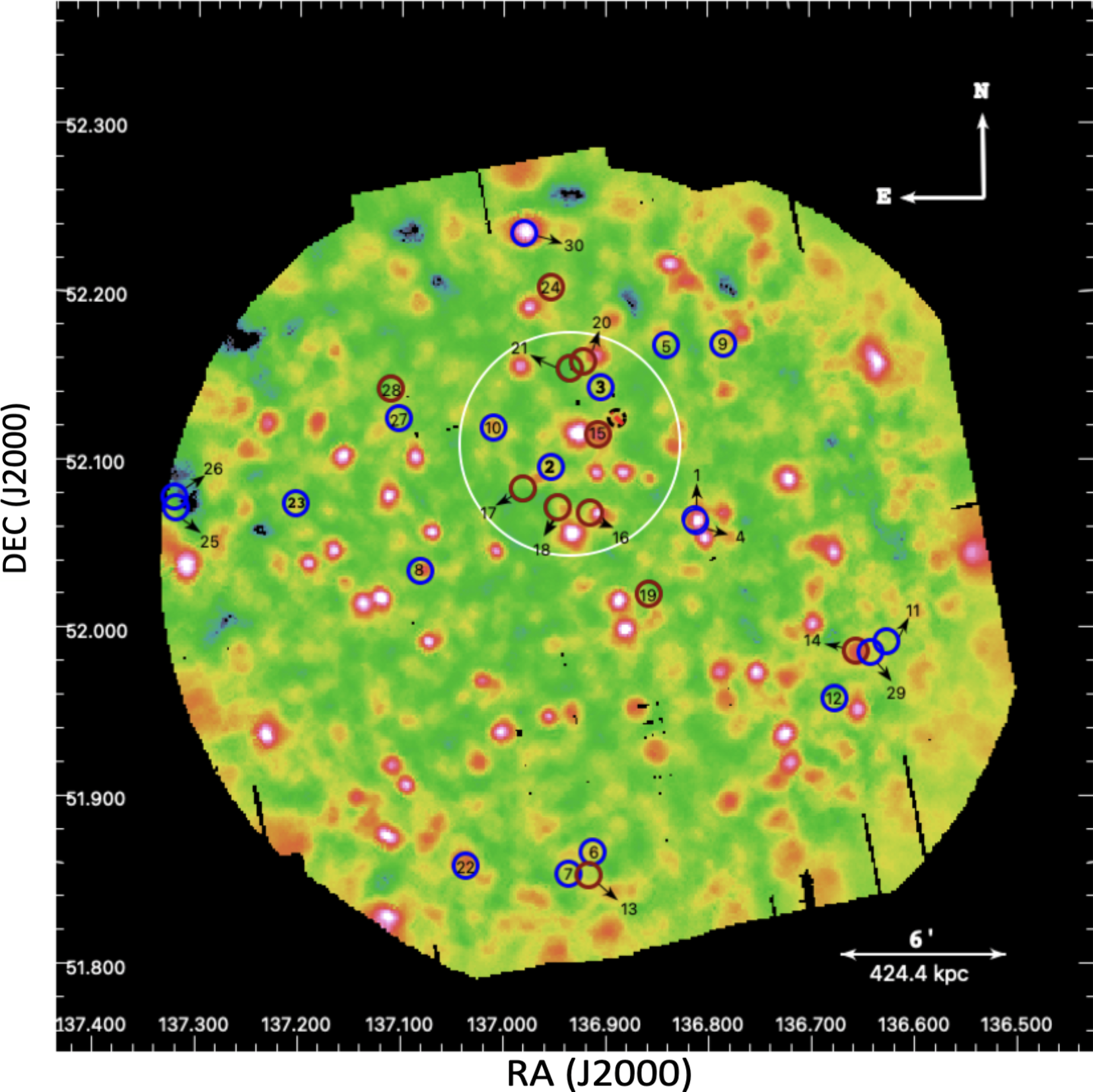}
\caption{0.5--2.0\,keV \xmm X-ray image of the galaxy cluster SDSS-C4\,3028, in units of cts/s/deg$^2$. Red and blue circles mark locations of the spectroscopically confirmed quiescent and SF member galaxies, respectively. The white circle represents the 4$^{\prime}$ (282.9\,kpc) core region around ID~15.
\label{fig:X-ray image}}
\end{figure}

\section{Introduction}
\label{sec:intro}

In the context of the $\Lambda$CDM model, the formation of galaxy clusters occurs through the gravitational collapse of initial density perturbations in the early cosmos, followed by hierarchical structure growth \citep[see][for a review]{2015SSRv..188...93P}. As the largest gravitationally bound structures in the Universe, galaxy clusters—particularly their mass functions—are valuable probes of cosmological parameters. Our primary knowledge of their masses are based on their X-ray emission from the hot ($\sim10^7-10^8$K), fully-ionized intracluster medium (ICM), which experiences gravitational heating due to the infall of matter into the dark matter potential. The ICM accounts for about 90\% of the total baryons in massive galaxy clusters. Observations show that their X-ray luminosity follows tight scaling relations with other cluster properties, such as mass and temperature, \citep[e.g.,][]{2009A&A...498..361P, 2006ApJ...640..691V}, indicating a degree of uniformity in the thermodynamic properties of galaxy clusters \citep[e.g.,][]{2019A&A...621A..41G}. However, substantial scatter is observed, particularly toward the X-ray faint end of the population. Furthermore, deviations from these scaling relations are more pronounced in dynamically disturbed or non-relaxed clusters, which may follow different scaling behaviors or display significantly larger intrinsic scatter compared to their relaxed counterparts \citep{2007ApJ...655...98N, 2007A&A...461..397P, 2022MNRAS.511.4991A}. The ICM undergoes heating from various non-gravitational sources, including active galactic nucleus (AGN) feedback and supernova-driven winds from star-forming (SF) galaxies. Such non-gravitational processes could elevate the ICM entropy and expel hot gas, making clusters more diffuse and under-luminous in X-rays. This presents a significant challenge in cluster selection and characterization, as X-ray observations are inherently biased toward detecting the brightest systems, whereas X-ray faint, but massive clusters are often missed. Studying X-ray faint/under-luminous clusters is therefore crucial to understanding the baryon physics in clusters and the use of clusters for cosmology. 

In addition to the ICM, galaxy clusters typically contain hundreds of galaxies whose properties and evolutionary paths are fundamentally shaped by the cluster environment. 
Multiple physical processes may be at play to affect the gas content of member galaxies and subsequently affect the fuel supply for star formation \citep[See][for review]{2007ApJ...671.1503M}. These processes can be broadly divided into two categories: 1) hydrodynamic processes, such as ram pressure stripping (RPS) and galaxy strangulation; 2) gravitational mechanisms, such as galaxy mergers and tidal interactions. In nearby mature galaxy clusters, the end result of quenching dominates, manifesting as a large fraction of elliptical galaxies having negligible star formation. In contrast, clusters at higher redshifts are found to display a larger fraction of blue galaxies, known as the Butcher-Oemler effect \citep{butcher1984evolution, 2013ApJ...775..126H}. Blue clusters are generally found to be significantly fainter in X-rays than clusters dominated by red galaxies. For instance, in an \xmm study of 43 blue galaxy clusters at $z=0.84$, \citet{Misato_2022} found that 14 were only marginally detected in the X-ray and no X-ray emission was detected in the rest. There is a long-standing debate in galaxy evolution over whether nearby cluster galaxies were quenched by internal factors \cite[e.g., stellar or AGN feedback;][]{10.1093/mnras/stae269} or by external environmental influences \citep[e.g., RPS;][]{2016A&A...596A..11B}. A nearby blue cluster would be one of the best ways to break the degeneracy between `nature' and `nurture'. 

One such blue cluster is SDSS-C4\,3028 at $z = 0.061$ (see Figure~\ref{fig:X-ray image}), which was initially detected in the spectroscopic sample of the Second Data Release (DR2) of the Sloan Digital Sky Survey \citep[SDSS;][]{Miller_2005}. SDSS-C4\,3028 was first noted by \citet{Hashimoto_2019} for its unusually high fraction of SF galaxies, $f_{\mathrm{SFG}}$ ($\sim~0.57$). To our knowledge, it is the nearest `blue' cluster. In this paper, we present new \xmm observations of SDSS-C4\,3028, complemented by a weak-lensing (WL) analysis using Subaru/Hyper Suprime-Cam (HSC) observations, as well as spectral energy distribution (SED) fitting of its member galaxies, with the goal of unveiling its dynamical state and evolutionary history. 

The paper is organized as follows: we report our observations and data reduction procedures, including member selection, spectral modeling, and analysis in Section~\ref{sec:Obs}. In Section~\ref{sec:Results}, we present the results of our X-ray and weak lensing analyses, cluster mass estimates, identification of AGN candidates, the likely BCG, and the star formation histories (SFHs) of member galaxies. In Section~\ref{sec:Discussion}, we discuss the implications of the absence of X-ray emission, focusing on the possible fate of the intracluster gas and the role of environmental mechanisms such as RPS in quenching of member galaxies. Finally, we summarize our main findings and conclusions in Section~\ref{sec:Conclusions}. We assume $H_0 = 70.0$\,km\,s$^{-1}$\,Mpc$^{-1}$, $\Omega_{\Lambda} = 0.73$, and $\Omega_{m} = 0.27$. At $z = 0.061$, this cosmology gives a luminosity distance of 273.7\,Mpc and an angular scale of 70.7\,kpc\,arcmin$^{-1}$. Unless otherwise stated, all uncertainty ranges are 68$\%$ confidence intervals (1$\sigma$).

\section{Observations and Data Reduction} \label{sec:Obs}

\subsection{Physical Properties of Member Galaxies}
\label{Galaxy Sample Properties}

With 12 SF galaxies out of 21 spectroscopically confirmed members, \cite{Hashimoto_2019} obtained a SF galaxy fraction of $f_{\rm SFG}=0.57$ for SDSS-C4\,3028, which is $>\,4\sigma$ larger than the median fraction of $\sim0.2$ for clusters at similar redshifts and virial masses \citep{2007A&A...461..411P}. We have added 9 more galaxies to the list, which are selected to have SDSS spectroscopically confirmed redshifts (\url{https://skyserver.sdss.org/dr18}) within 3$\sigma$ of the median redshift ($z=0.061$), to be within the field-of-view (fov) of the \xmm pointing, and to have $r$-band absolute magnitudes $M_{r}$ brighter than $-20.1$. The $M_{r}$ cut is selected to ensure the completeness of SDSS spectroscopic observations and to maintain consistency with the previous study by \citet{Hashimoto_2019}. This results in a total of 19 SF and 11 quiescent (QS) members, classified based on their stellar mass and star-formation rate (SFR), using the same criteria as \citet{Hashimoto_2019} (see Figure 4 of \citet{Hashimoto_2019}), corresponding to a $f_{\rm SFG}$ of $0.63 \pm 0.08$. Notably, all newly identified galaxies are located beyond 300\,kpc from the cluster center (later defined as outskirts). This difference may be attributed to the friends-of-friends method employed in \citet{Hashimoto_2019}, which tends to miss member galaxies residing on the outskirts of the cluster \citep{2020RAA....20..207W}.

Table~\ref{tab:Properties} lists the properties of all 30 member galaxies with columns containing the ID number assigned to a galaxy in this work, right ascension (J2000), declination (J2000), spectroscopic redshift (from SDSS), 0.5--2.0\,keV and 2.0--10.0\,keV X-ray luminosities, stellar mass, SFR, SF or QS classification, and morphological category. SFRs and stellar masses are publicly available in the SDSS catalog\footnote[1]{\url{https://wwwmpa.mpa-garching.mpg.de/SDSS/}} \citep{2003MNRAS.346.1055K, 2004MNRAS.351.1151B, 2004ApJ...613..898T, 2007ApJS..173..267S}. The X-ray measurements are presented in Section~\ref{sec:Members}. Finally, the galaxy morphologies are taken from the NED database\footnote[2]{\url{https://ned.ipac.caltech.edu/}}. 

%We also use $K$-band infrared imaging from the 2MASS public archive to determine the $K$-band luminosity (tracing the stellar light distribution) of the member galaxies. We adopt annular regions based on the surface brightness profiles. The local infrared background surface brightness,  determined using an offset region is subtracted from each region. We convert source counts in each region to a corresponding magnitude and correct for Galactic extinction. We adopt a Solar $K$-band absolute magnitude of M$_{K_{\odot}} = 3.28$ \citep{1998gaas.book.....B}.

\subsection{\xmm}
 
SDSS-C4\,3028 was observed with the \xmm European Photon Imaging Camera (EPIC) for 58\,ksec on November 10--11, 2021 (Observation ID: 0883100101, 0883100201). The EPIC pointing, aimed at (RA, DEC) = (09h 07m 47.61s +52d 02$^{\prime}$ 03.5$^{\prime\prime}$), covers the entire cluster, extending to $\sim14^{\prime}$, in all directions. The three EPIC instruments, MOS1, MOS2, and pn, were operated in full-frame mode. We use version xmmsas-v20.0.0 of the \xmm Science Analysis System (SAS)\footnote[3]{\url{https://www.cosmos.esa.int/web/xmm-newton/sas-download}} to perform data reduction and analysis. All ODF files are processed using {\tt emchain} and {\tt epchain} to generate the event files with the latest calibrations. To remove soft flares from the MOS and pn data, we utilize the XMM-ESAS tools {\tt mos-filter} and {\tt pn-filter}, respectively \citep{snowden2014cookbook}. For MOS data, we only include events with FLAG $ == 0$ and PATTERN\,$ <= 12$, while for pn data, we consider events with FLAG\,$ == 0$ and PATTERN\,$ <= 4$. After applying filters to remove background flares, the effective exposure times are 22\,ksec, 26\,ksec, and 16\,ksec for MOS1, MOS2, and pn, respectively. We employ the XMM-ESAS routine {\tt cheese} to detect point sources, down to $1 \times 10^{-14}$\,erg\,s$^{-1}$\,cm$^{-2}$ in the 0.4--7.2\,keV band. We remove the out-of-time pn events for both the spectral and imaging analyses. Spectra and images are extracted using the routines {\tt mos-spectra} and {\tt pn-spectra}. The resulting data, along with the XMM-ESAS CalDB files describing the quiescent particle background (QPB), are then used by {\tt mos-back} and {\tt pn-back} to create images and spectra of the QPB. We use the SAS task {\tt rmfgen} to generate redistribution matrix files (RMFs), and the task {\tt arfgen} to produce ancillary response files (ARFs) for the region and each of the detectors. We combine individual source and background images produced by XMM-ESAS tools using the {\tt comb} routine. Finally, the {\tt adapt} routine is used to adaptively smooth the combined image. The final background-subtracted, vignetting and exposure-corrected, smoothed, 0.5--2.0\,keV image, in units of cts/s/deg$^2$, is shown in Figure~\ref{fig:X-ray image}.

\subsubsection{Background Modeling} \label{bg_modeling}

We account for two distinct background components: the astrophysical X-ray background (AXB) and the non-X-ray background (NXB). 

The AXB model comprises the cosmic X-ray background \citep[CXB;][]{de20042}, for which we employ a power law model, denoted as {\tt pow}$_{\mathrm{CXB}}$, with a fixed photon index of $\Gamma = 1.41$, the Milky Way (MW) emission modeled with a thermal emission model, {\tt apec}$_{\mathrm{MW}}$, with a temperature of 0.2\,keV \citep{McCammon_2002}, and the Local Bubble (LB) emission modeled with a thermal emission model,  {\tt apec}$_{\mathrm{LB}}$, with a temperature of 0.12\,keV. We fix the metal abundances to 1 Z$_{\odot}$ and the redshifts to 0, for {\tt apec}$_{\mathrm{MW}}$ and {\tt apec}$_{\mathrm{LB}}$. Furthermore, the photoelectric absorption ({\tt phabs}) model characterizes the foreground (Galactic) cooler gas that is expected to absorb all the AXB components except {\tt apec}$_{\mathrm{LB}}$. We adopt a Galactic hydrogen column of $N_{\mathrm{H}} = 2 \times {10^{20}}$ cm$^{-2}$ \citep{1990ARA&A..28..215D}. We utilize spectra extracted from RASS over an annular region with an inner radius of $0.5^\circ$ and an outer radius of $1^\circ$ to help constrain the AXB.

The NXB component consists of a continuum spectrum and a set of fluorescent instrumental lines for each MOS and pn detector. A set of Gaussian lines are used to model the fluorescent instrumental lines produced by the hard particles. The centroid energies of the Gaussian lines are listed in Table~2 in \cite{2017ApJ...851...69S} and we set an upper limit of 0.3\,keV on the line width. To model the continuum particle background component, we adopt a broken power law model, {\tt bknpow}, with the energy break fixed at 3\,keV. 

\begin{figure}[!t]
\centering
\includegraphics[width=\linewidth]{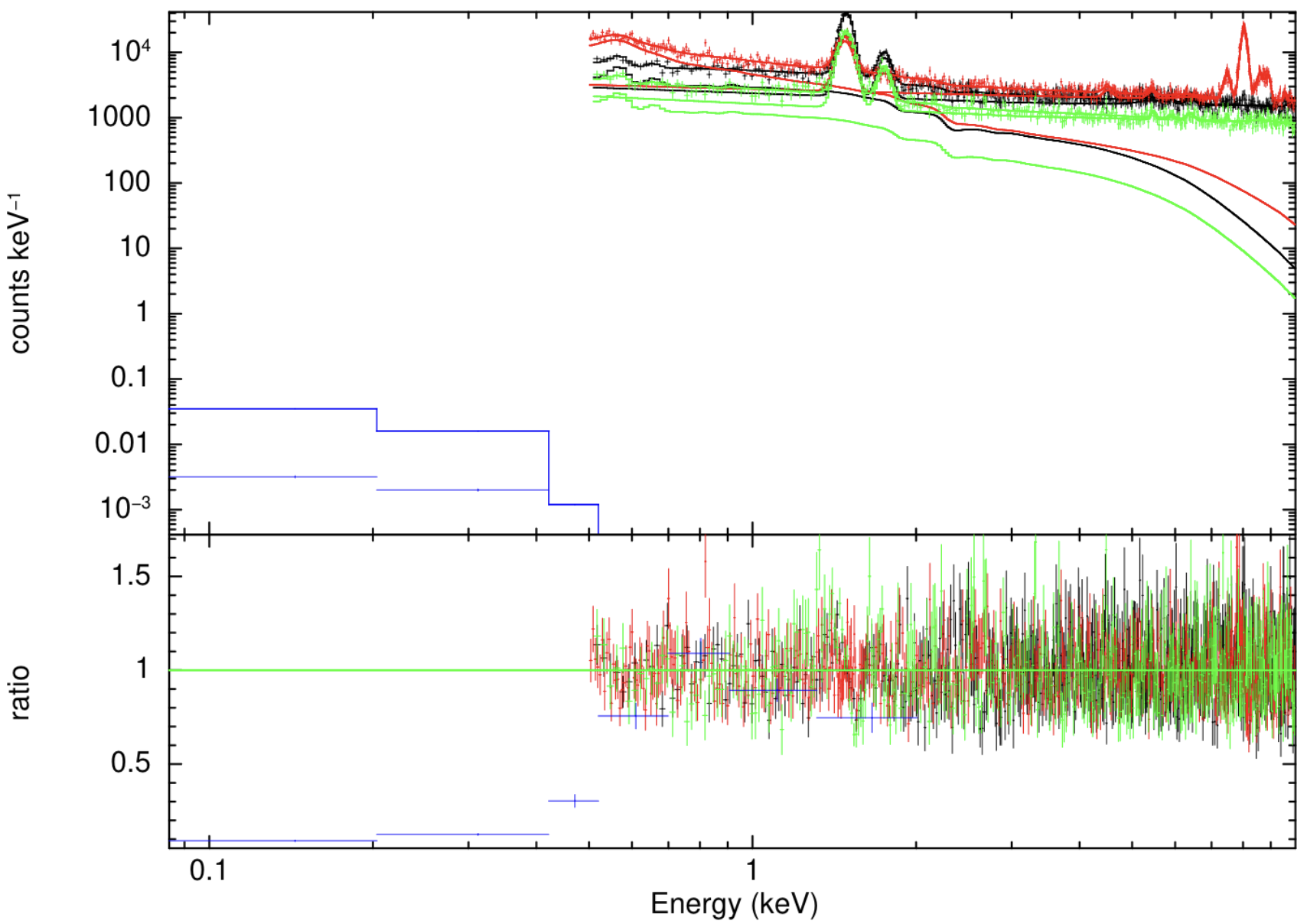}
\caption{MOS1 (green), MOS2 (black), and pn (red) spectra are extracted from a circle of radius 11.6$^{\prime}$ (820.6\,kpc) around the SDSS-C4\,3028 observation. The RASS (blue) spectrum is taken from an annular region with inner and outer radii as 0.5$^{\circ}$ and 1$^{\circ}$, respectively. The upper panel shows the fit with the model background subtracted and the lower panel plots the ratio of the data to the model. The best-fit ICM component is negotiable and consistent with 0.}
\label{fig:Spectrum}
\end{figure}

\subsubsection{Spectral Analysis} \label{spectral_analysis}

To simultaneously determine the level of the ICM and the background, we extract spectra from a circular region with a radius of 11.6$^{\prime}$ (820.6\,kpc) while masking the member galaxies and point sources. We mask all member galaxies with a circular region with a radius of R$=0.6$--$1.5^{\prime}$ (42.4--106.1\,kpc). In addition, point sources are masked with a circular region of R$=0.3^{\prime}$ (21.2\,kpc). We perform the fitting with {\tt XSPEC-12.12.0} and using the C-statistic. We limit the spectral analysis to the energy ranges of 0.5--10.0\,keV for both MOS and pn spectra. Each of the observations are fit jointly to two sets of models. The first model set takes the form of {\tt apec}$_{\mathrm{LB}}$ + {\tt phabs} × ({\tt apec}$_{\mathrm{ICM}}$ + {\tt apec}$_{\mathrm{MW}}$ + {\tt pow}$_{\mathrm{CXB}}$). The model consists of the X-ray emission from the ICM ({\tt apec}$_{\mathrm{ICM}}$) and the AXB sources as presented above. We fix the metal abundance, redshift and temperature to 0.2\,Z$_{\odot}$, 0.061, and 2\,keV, respectively, for the ICM thermal model. We adopt the solar abundances in \citet{2009ARA&A..47..481A} for the spectral analysis. The second model set describes the NXB components. The spectra and best-fit model are shown in Figure~\ref{fig:Spectrum}. As listed in Table~\ref{tab:Norm}, the levels of the three AXB components are fully consistent with the expected sky background levels \citep{yoshino2009energy}. The best-fit ICM component is negotiable and consistent with 0. 

\begin{deluxetable}{lcccc}
\tabletypesize{\scriptsize}
\tablewidth{0pt} 
\tablecaption{X-ray background determinations \label{tab:Norm}}
\tablehead{
\colhead{Parameters} & \colhead{CXB} & \colhead{MW} & \colhead{LB} & \colhead{C/d.o.f}
}
\startdata
\textit{kT} or $\Gamma$ & 1.41 & 0.2 & 0.12 & \\
Norm$^\mathrm{a}$  & 13.72$^{+2.15}_{-2.29}$ & 7.21 $^{+1.22}_{-1.26}$ & 18.61$^{+2.52}_{-2.52}$ & 1778.77/1412 \\
SB$^\mathrm{b}$  & 3.13$^{+0.49}_{-0.52}$ & 0.97 $^{+0.16}_{-0.17}$ & 0.73$^{+0.10}_{-0.10}$ & 
\enddata
\tablenotetext{a}{Model parameters for \xmm observations: normalization per steradian.}
\tablenotetext{b}{Surface brightness in 0.5–2.0 keV, in units of $10^{-8}$ erg s$^{-1}$ cm$^{-2}$ sr$^{-1}$.}
\end{deluxetable}

\begin{figure*}[h!]
\centering
\includegraphics[width=\linewidth]{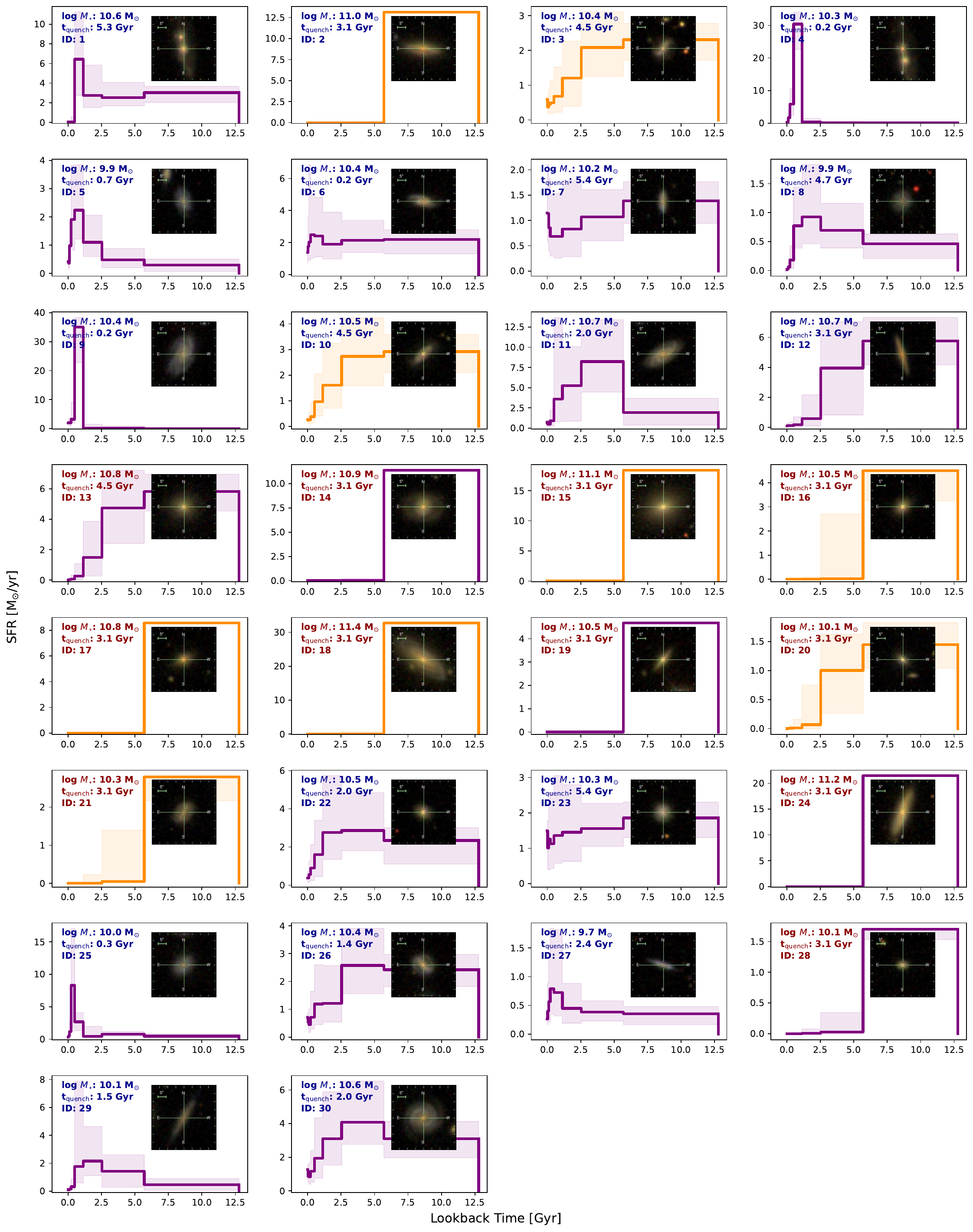}
\caption{SFHs of the member galaxies of SDSS-C4\,3028. Blue and red inserted texts represent star-forming and quiescent galaxies, respectively. The orange and purple plots represent the galaxies within the core and the outskirts, respectively. The shaded regions represent the 16$^{\rm th}$- 84$^{\rm th}$ percentiles. Every galaxy is labeled with its ID, stellar mass and quenching timescale. The insets show the corresponding SDSS DR18 optical images.} 
\label{fig:Members}
\end{figure*}

\subsection{Subaru weak-lensing data}\label{sec:Weak lensing}

SDSS-C4\,3028 was observed in January 2023 using the HSC as part of the Abell 746 field campaign \citep[PI: H. Cho;][]{2024ApJ...962..100H}. These observations were complicated by the presence of the bright foreground star f UMa ($V=4.48$) near the field center, which generated significant optical ghosts \citep[e.g.,][]{2022PASJ...74..247A} with the most severe effects at the field periphery. As SDSS-C4\,3028 unfortunately lies in the periphery of the HSC field, we implement manual masking of affected regions through visual inspection of individual exposures prior to shape measurement. Specifically, we mask optical ghosts and artifacts frame by frame near the cluster region. However, we note that the result should be treated with caution, since any unmasked faint or diffuse artifacts can harm the shape measurement. Subsequent WL analysis utilizes the cleaned $r$-band imaging, with source selection employing photometric ($22<r<26$) and shape quality cuts as described in \citet{2024ApJ...962..100H}. No color criterion is applied because the majority of galaxies observed in this low-redshift cluster field are background sources, minimizing contamination from cluster members \citep{2024NatAs...8..377H, 2025ApJS..277...28F}. The final source density is 20 arcmin$^{-2}$.

\subsection{Star-formation histories}
We use the SDSS ({\tt ugriz}), 2MASS ({\tt JHK}), \textit{WISE} ({\tt W1 W2 W3 W4}) photometry and spectra from SDSS for fitting spectral energy distributions (SEDs). We obtain multi-wavelength photometry for each galaxy by querying SDSS DR18, 2MASS, and \textit{WISE} using \texttt{astroquery}. SDSS optical magnitudes\footnote[4]{\url{https://www.sdss4.org/dr17/algorithms/magnitudes/}} ({\tt ugriz}) are retrieved as composite model magnitudes; near-infrared magnitudes ({\tt JHK}) are obtained from 2MASS Extended or Point Source Catalogs with Vega-to-AB conversions applied; and mid-infrared \textit{WISE} magnitudes ({\tt W1 W2 W3 W4}) are extracted from the AllWISE catalog and similarly converted to AB magnitudes. All photometric errors are propagated appropriately, and magnitudes are converted to maggies (linear flux units) for use in SED fitting. The insets in Figure~\ref{fig:Members} show colored cutouts of member galaxies taken from the SDSS DR18 catalog\footnote[5]{\url{http://cas.sdss.org/dr18/VisualTools/navi}}. 

We employ \texttt{Prospector}, a Bayesian inference SED fitting code, using the sampling code \texttt{dynesty} \citep{2020MNRAS.493.3132S}, which adopts a nested sampling procedure and simultaneously provides estimates of the Bayesian evidence and posterior \citep{2021ApJS..254...22J}. For the basic setup of \texttt{Prospector}, we employ the Flexible Stellar Population Synthesis (\texttt{FSPS}) code \citep{2009ApJ...699..486C} for stellar population synthesis, where the Modules for Experiments in Stellar Astrophysics (\texttt{MESA}) Isochrones and Stellar Tracks (\texttt{MIST}) are used \citep{2016ApJ...823..102C, 2016ApJS..222....8D}. We fix the redshift to the spectroscopic redshift from SDSS. We adopt a single stellar metallicity that is varied with a prior that is uniform in $\rm \log(Z_{\ast}/Z_{\odot})$
between – 1.0 and 0.19, where Z$_{\odot}=0.0142$. We assume the \citet{2000ApJ...533..682C} dust attenuation law and fit the V-band dust optical depth with a uniform prior, $\tau_V \in (0.001, 5)$. We choose the \citet{2003PASP..115..763C} initial mass function (IMF) with lower and upper limits of 0.1\,M$_{\odot}$ and 300\,M$_{\odot}$, respectively.

We adopt a similar \texttt{Prospector} model as adopted in \citet{2017ApJ...837..170L, 2019ApJ...876....3L} \citep[also see][]{2024MNRAS.527.3291J}. We run \texttt{Prospector} based on a flexible model, for which we assume that the star-formation history (SFH) is step-wise constant in 8 time bins. We fit for the ratio of the SFR between the time bins (7 free parameters) plus the total stellar mass. We use the standard continuity prior \citep{2019ApJ...876....3L} and assume a uniform prior for the stellar mass in the range of $10^9$ to $10^{12}$\,M$_{\odot}$. The first two-time bins are fixed to $0-30$\,Myr and $30-100$\,Myr, while the remaining bins are logarithmically varied in time. 

\section{Results} \label{sec:Results}
\subsection{Structural analysis of the galaxy cluster}
\label{Structural Analysis}
\begin{figure}[ht]
\centering
\includegraphics[width=\columnwidth]{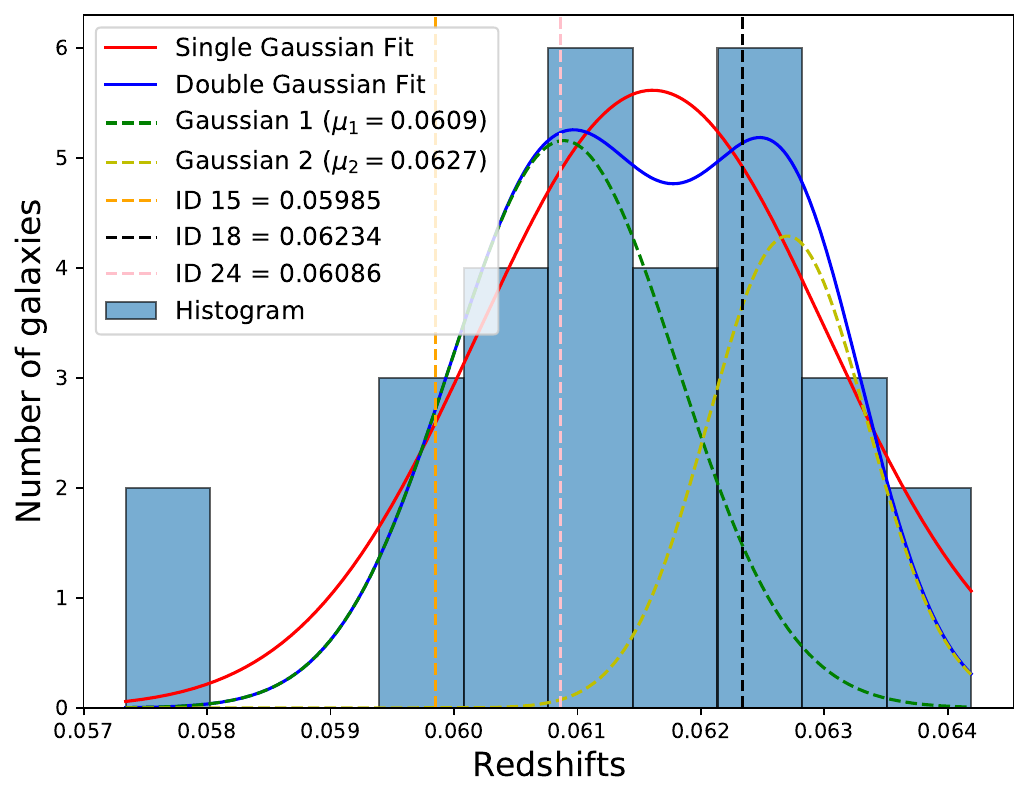}
\caption{The histogram illustrates the redshift distribution of member galaxies. The solid red curve corresponds to a single Gaussian fit. The solid blue curve represents a double Gaussian fit, with its two components depicted by dashed green and yellow curves. The vertical dashed lines mark three massive galaxy members: orange (ID~15), pink (ID~24), and black (ID~18). The redshift distribution is well fit with both the single and double Gaussian models, leaving the presence of a bimodal distribution—suggesting two potential sub-clusters within the system—and the occurrence of a merger inconclusive.}
\label{fig:z_bimodal}
\end{figure}

We present the redshift distribution of all 30 members in Figure~\ref{fig:z_bimodal} together with the best fitting single and double Gaussian models.
The single Gaussian model yields a mean redshift of $z_{\text{cluster}} = 0.061$, representing the overall cluster redshift under the assumption of a uni-modal distribution. The double Gaussian model, which suggests a possible bimodal structure, yields two redshift peaks at $z_1 = 0.061$ and $z_2 = 0.063$, which may correspond to two dynamically distinct substructures or merging components within the cluster. To ensure robustness against binning choices, we randomly selected the number of bins between 6 and 20 and repeated the process 300 times. We use the Bayesian Information  Criterion \citep[BIC;][]{1978AnSta...6..461S} to choose between models with different number of Gaussian components. The model that minimizes the BIC is the model that is most supported by the data. We find that the distribution does not strongly favor either the double Gaussian model or the single Gaussian model, with probabilities of 0.55 and 0.45, respectively, depending on the number of bins. Therefore, it is unclear whether SDSS-C4\,3028 consists of a single cluster, or two sub-clusters/groups that are in the process of merging. In the merging scenario, we observe no distinct separation in the spatial distributions of the two groups. In that case, the merger event is likely to be along the line-of-sight (LOS). However, the number of member galaxies is small, making the analysis susceptible to low-number statistics. Further spectroscopic redshift follow-up observations over a larger fov would be valuable for better constraining the dynamical structure of the cluster. 

\begin{figure*}[ht!]
\centering
\includegraphics[width=\linewidth]{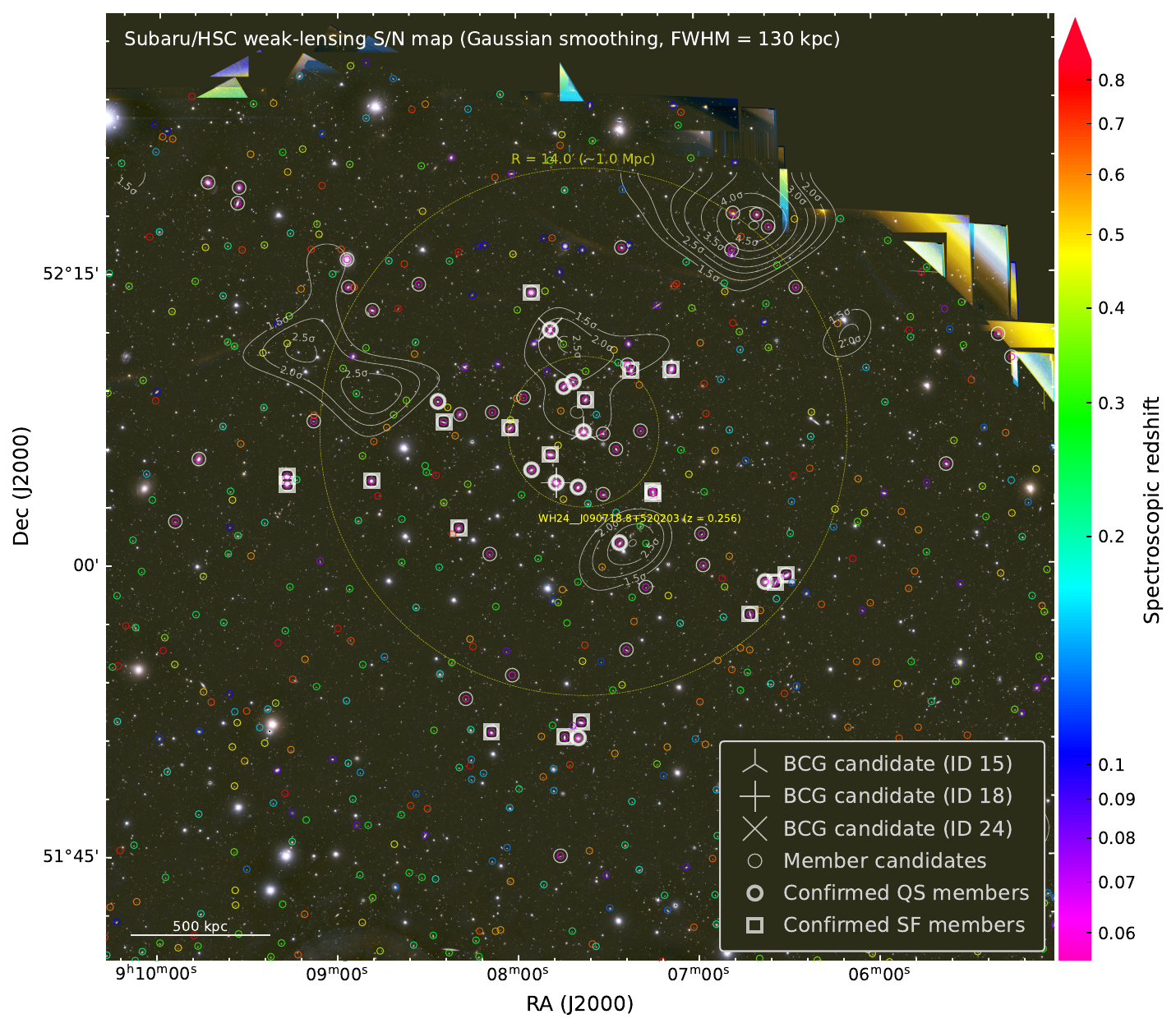}
\caption{A zoomed-in view of a $\sim50 \times 50$ arcmin$^2$ region centered on SDSS-C4\,3028, extracted from the color-composite image of the Subaru/HSC field centered on A746. White contours represent the WL significance map, starting from 1.0$\sigma$ with 0.5$\sigma$ linear intervals. Small circles, color-coded according to the color bar at the right, indicate galaxies with spec-$z$ from SDSS and Dark Energy Spectroscopic Instrument (DESI) DR1 \citep{2025arXiv250314745D}. 
Large white symbols mark the BCG from \citet{Hashimoto_2019} (ID~18; $+$), the BCG identified in HSC imaging (ID~24; $\times$), BCG candidate near the peak of member galaxies' density (ID~15; upside-down Y), and galaxies within a redshift range similar to that of the cluster members identified in \citet{Hashimoto_2019} (circles). QS and SF members from Table~\ref{tab:Properties} are marked by thicker circles and squares, respectively. The big dotted circle represents 14$^{\prime}$ radius from the central BCG (ID~15), which is similar in size to the estimated projected $R_{200}$. A 3$\sigma$ WL mass clump to the south is associated with a galaxy cluster at $z = 0.256$ \citep[e.g.,][labeled `WH24\_J090718.8+520203' in the image]{2024ApJS..272...39W}. In addition, by utilizing SDSS and DESI $z_{\rm spec}$ information, we find that the peak of this clump spatially coincides with two galaxies (marked by green circles) with $z_{\text{north}} = 0.256$ and $z_{\text{south}} = 0.254$, supporting the interpretation that the mass clump corresponds to this background galaxy cluster.}
\label{fig:Weak_lensing}
\end{figure*}

\begin{figure*}[ht!]
\centering
\includegraphics[width=\linewidth]{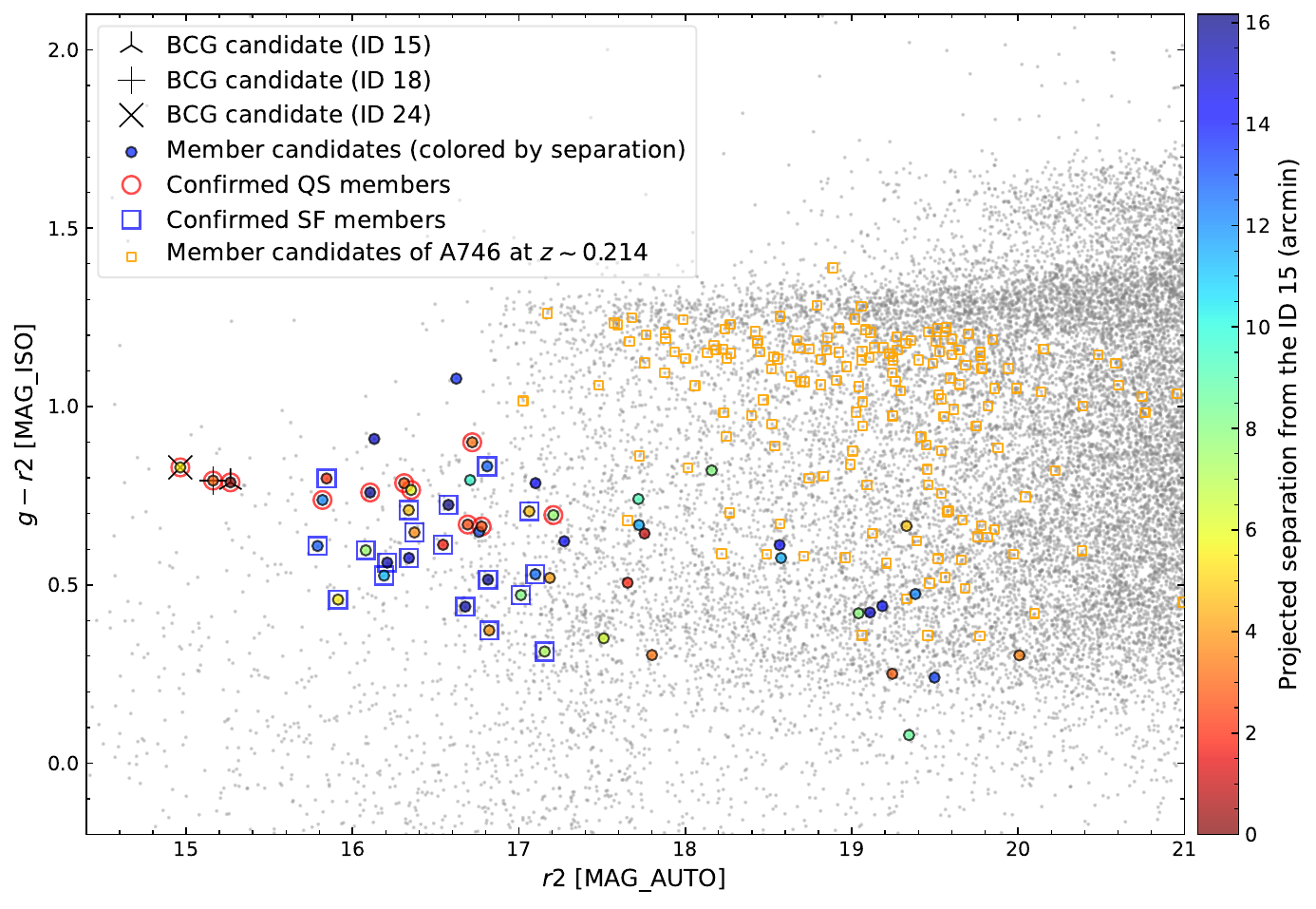}
\caption{Color-magnitude diagram (CMD) of the Subaru/HSC A746 field \citep[See][for details]{2024ApJ...962..100H}, combining photometric data with spectroscopic redshifts from SDSS and DESI DR1. 
Gray points represent all objects detected in the Subaru/HSC field. QS (red circles), SF (blue squares), and the BCG identified by \citealt{Hashimoto_2019} (ID~18; black $+$ sign), as well as the BCG detected in Subaru/HSC observations (ID~24; black $\times$) and central BCG (ID~15; upside-down Y), are shown. SDSS-C4\,3028 member candidates within a 14$^{\prime}$ radius from ID~15 ($R_{\text{ID15}}$), which is similar in size to the estimated projected $R_{200}$, are color-coded according to their projected separation. This separation allows for assessing cluster-centric concentration by indicating how far QS, SF, or bluer member candidates are located from the ID~15. Orange squares indicate member candidates of the HSC-field central cluster A746 at $z = 0.214$.}
\label{fig:Weak_lensing_CMD}
\end{figure*}

\subsection{Identifying BCG}\label{sec:Identifying BCG}
Among the 30 member galaxies, 9 reside within a `core' region of radius R$<280$\,kpc ($\sim4^{\prime}$) from the cluster center (assumed to be ID~15), while the remaining galaxies are scattered throughout the cluster `outskirts'. The three most massive member galaxies (IDs 15, 18 and 24), are the BCG candidates of SDSS-C4\,3028. While the BCG is literally defined as the most optically luminous galaxy in a cluster, in practice, its identification can be ambiguous, particularly in dynamically young or complex systems. For example, in the Virgo cluster, M49 is the most optically luminous galaxy (especially in the $r$-band) and thus the formal BCG \citep{2019AJ....158....6S}, while M87, located at the cluster center and coinciding with the X-ray peak, is often considered the central dominant galaxy of Virgo \citep{1994Natur.368..828B}. 
However, it is difficult to identify BCG using X-ray peak for X-ray faint clusters. In SDSS-C4\,3028, the redshifts of the more massive and less massive sub-clusters, identified by the double-Gaussian model, are centered on ID~24 and ID~18, respectively. Our HSC imaging analysis identifies the BCG as ID~24 (see Section~\ref{sec:Weak lensing}). ID~15 and ID~18 reside within the core, while ID~24 resides in the outskirts. ID~15 is the only elliptical galaxy of the three and closest to peak of the galaxy density. ID~15 and ID~24 contains diffuse X-ray emission, while ID~18 lacks a hot ISM. Overall, SDSS-C4 3028 contains multiple BCG-like galaxies, making it difficult to identify a single dominant BCG. As in the Coma cluster, the presence of multiple BCGs suggests that the cluster is in a dynamically active state, likely the result of a recent merger or accretion event \citep{2007A&A...468..815G}. 

\subsection{Weak-lensing analysis}

Figure~\ref{fig:Weak_lensing} shows a zoomed-in view of the color-composite image of the Subaru/HSC field, centered on SDSS-C4\,3028. From the mass map and $z_{\rm spec}$ catalog (SDSS and DESI) data, we identified a $\sim2.5\sigma$ clump at the location of SDSS-C4\,3028 and a $\sim3\sigma$ clump to the south, which is associated with a galaxy cluster at $z\sim0.256$. For the SDSS-C4\,3028, ID~24 (marked with $\times$) is near the peak of the $\sim2.5\sigma$ clump. Additionally, there appears to be $\sim5\sigma$ detection of a clump to the northwest of the SDSS-C4\,3028. Spectroscopic redshift information reveals a notable overlapping concentration of galaxies with redshifts similar to that of SDSS-C4\,3028, suggesting that the northwestern clump could represent a filamentary structure connected to the main cluster or a subhalo within the larger system. However, the proximity of this feature to the CCD boundary of the HSC mosaic, where point-spread function (PSF) anisotropies are known to introduce spurious alignments, raises concern about the reliability of the detection. PSF elongation near field boundaries can mimic a WL signal. Given these systematics and the lack of independent evidence for a distinct structure at that location, we interpret the $\sim5\sigma$ signal with caution. A future observation with deeper exposure and having the SDSS-C4\,3028 centered in the observation field can help obtain a more stringent constraint on its mass as well as the presence of other potential cluster candidates that could be connected via filaments. 

In order to explore the spatial distribution of the galaxies around the $\sim2.5\sigma$ clump, Figure~\ref{fig:Weak_lensing_CMD} plots the color-magnitude diagram (CMD) of the identified cluster members within a 14$^{\prime}$ radius (projected $R_{200}$) from ID~15 identified as probable BCG nearest to the cluster center, color-coded according to their projected separation. This separation allows for assessing cluster-centric concentration by indicating how far QS, or SF, or fainter member candidates are located from ID~15.

\subsection{Cluster mass estimates}\label{sec:mass}

We measure the LOS velocity dispersion ($\sigma_{\rm p}$), projected virial radius ($R_p$), and halo mass ($M_{\rm halo}$) of the galaxy cluster using \citep{yang2007galaxy, 2007ApJ...655..790C, 2013MNRAS.430.2638M}:
\begin{equation}
    \sigma_{\rm p}^2 = \frac{\sum_i (V_i - V_c)^2}{N_{\rm member}-1},
\end{equation}
\begin{equation}\label{eq:virial_radius}
    R_p = \frac{N_{\rm member}(N_{\rm member} -1)}{\sum_{i>j}{R}_{ij}^{-1}},
\end{equation}
\begin{equation}
   M_{\rm halo} = 10^{15} \left(\frac{\sigma_{\rm p}}{A_{\rm 1D}}\right)^{\frac{1}{\alpha}}\frac{1}{h(z)}
   % M_{\rm halo} = \frac{3\pi}{2} \frac{\sigma_{\rm p}^2 R_p}{\rm G}
\end{equation}
where $A_{\rm 1D}$ and $\alpha$ are the constants from \citet{2013MNRAS.430.2638M}, $V_c$ is the mean of the recessional velocities of member galaxies, $V_i$. The number of member galaxies is denoted by $N_{\rm member}$. $R_{ij}$ is the projected distance between $i$th and $j$th member galaxies.

The LOS velocity dispersion of $\sim462^{+58}_{-74}$\,km/s and projected virial radius of $\sim0.81$\,Mpc correspond to a dark matter halo mass of $M_{\rm halo}=1.03\pm{0.40}\times 10^{14}$\,M$_{\odot}$. The errors are calculated using bootstrapping. 

From the Subaru/HSC WL analysis, we estimate the mass of SDSS-C4\,3028 by fitting a Navarro–Frenk–White \citep[NFW;][]{1996ApJ...462..563N, 1997ApJ...490..493N} profile. We fix the mass-concentration relation using \citet{2021MNRAS.506.4210I} and centered the profile at the mass peak which corresponds to the BCG (ID~24) identified in the HSC imaging. We estimate the total mass of the cluster to be $M_{200}=1.3\pm0.9\times10^{14}$~M$_{\odot}$ (reduced $\chi^2 = 1.1$). We obtain a consistent result after taking into account the neighboring cluster at $z=0.256$ by fitting multi-halo NFW models \citep[e.g.,][]{2025NatAs.tmp...93H}.

We also calculate the halo mass using the red-sequence galaxy cluster richness estimate, following \citet{andreon2015making}. Since SDSS-C4\,3028 is an unusual blue cluster and to ensure the richness-based estimate is not biased by its lack of red-sequence galaxies, we assume $f_{\rm SFG}=0.2$, similar to typical clusters, such that 80\% of the member galaxies are red sequence galaxies. The resulting halo mass is  $1.14\pm0.20\times 10^{14} {\rm M}_{\odot}$. In addition, we estimate the halo mass from total $r$-band luminosity of member galaxies, following \citet{2013AJ....146..124H}. This gives a halo mass of $2.45^{+0.34}_{-0.32}\times 10^{14} {\rm M}_{\odot}$. The different halo mass estimates are summarized in Table~\ref{tab:M_halo}. 

\begin{deluxetable}{lcccc}
\tabletypesize{\scriptsize}
\tablewidth{0pt} 
\tablecaption{Cluster mass determinations \label{tab:M_halo}}
\tablehead{
\colhead{\textbf{Estimation method}} & \colhead{\textbf{$M_{\text{halo}}$($\times 10^{14} \rm M_{\odot}$)}}
}
\startdata
LOS velocity dispersion$^a$ & 1.03$^{+0.40}_{-0.40}$\\
WL analysis$^b$ & 1.30$^{+0.90}_{-0.90}$\\ 
Optical Richness$^c$ & 1.14$^{+0.20}_{-0.20}$\\
Total $r$-band luminosity$^d$ & 2.45$^{+0.34}_{-0.32}$\\
\enddata
\tablenotetext{a}{Section~\ref{sec:mass}}
\tablenotetext{b}{Section~\ref{sec:Weak lensing}}
\tablenotetext{c}{\citet{andreon2015making}}
\tablenotetext{d}{\citet{2013AJ....146..124H}}
\end{deluxetable}

%\begin{figure*}
%    \includegraphics[width=1.0\textwidth]{Figures/SB_LxLk_Combined.pdf}
%    \caption{\textit{Left:} X-ray surface brightness profiles, centered on four cluster members with diffuse X-ray emission detected. The numbers in blue and red represent the IDs assigned to the star-forming and quiescent galaxies in this work, respectively. The horizontal pink line marks the average local background level. The PSF (dashed black line) denotes the point spread function profile of a nearby point source. \textit{Right:} X-ray luminosity as a function of $K$-band luminosity for field, cluster/group, and group-center (NGC\,5813, NGC\,5846, and NGC\,4636) galaxies studied by \citet{2015ApJ...806..156S}. Overlaid are ID~15 (red star) and ID~24 (grey star). ID~15 is near the top of the observed scatter, suggesting that its luminosity ratio is comparable to that of the central galaxies of nearby groups.
%    \label{fig:SB}}
%\end{figure*}

\subsection{Absence of extended ICM X-ray emission} \label{sec:ab_ICM}

A visual inspection of the X-ray image (Figure~\ref{fig:X-ray image}) reveals that the diffuse emission is uniform, with no obvious X-ray brightness enhancement associated with the cluster, suggesting a lack of ICM. This uniformity further suggests a background/foreground nature of the emission. The spectral analysis, as shown in Figure~\ref{fig:Spectrum} and listed in Table~\ref{tab:Norm}, confirms the absence of ICM (see Sections~\ref{bg_modeling} and \ref{spectral_analysis} for details). We put a $3\sigma$ upper limit of $L_{\rm X}$ (0.1--2.4\,keV) $=7.7 \times 10^{42}$ erg s$^{-1}$ on its X-ray luminosity. %This is nearly two orders of magnitude below the typical X-ray luminosity predicted for clusters at similar redshift, based on established scaling relations as presented in \citet{2009A&A...499..357G}.
As shown in Table~\ref{tab:M_halo}, the virial mass of SDSS-C4\,3028 is likely ranging from $\sim1-3\times10^{14}\, \rm M_\odot$, rivaling that of the Virgo cluster. Given its velocity dispersion of $\sim462$ km/s, this cluster is intermediate in mass between Virgo, which has a velocity dispersion of $\sim600$ km/s \citep{2020A&A...635A.135K}, and Fornax, with a velocity dispersion of $\sim350$ km/s \citep{2022A&A...657A..93C}.  Its $3\sigma$ upper limit of $L_{\rm X}$ is below what would be expected from the $L_{\rm X}$--$M_{\rm halo}$ relation of \citet{Wang_2014}.% by a factor of 0.1 -- 6.  

The absence of enhanced X-ray brightness in the cluster could be a consequence of low intracluster gas density or of the gas temperature being well below the cluster's virial temperature \citep{2011A&A...530A..27C, 2022A&A...666A..22R}. To quantify this, we employ a similar approach to that used for high redshift ($z=0.84$) blue clusters reported in \citet{Misato_2022}, which have comparable weak-lensing masses and X-ray luminosities. We estimate the gas mass ($M_{\text{gas}}$) by first determining the electron density using the best-fit normalization of the {\tt apec}$_{\mathrm{ICM}}$ model. The normalization is given by 
\begin{equation}
\text{norm}=\frac{10^{-14}}{4\pi [D_A(1+z)]^2}\int n_e n_H dV,
\end{equation}
where $D_A$ is the angular diameter distance to the source (cm), $dV$ is the volume element (cm$^3$); $n_e$ and $n_H$ are the electron and hydrogen densities (cm$^{-3}$), respectively assuming an ionized plasma $n_H = n_e/1.2$.
To estimate the total gas mass, we integrate the electron density over a spherical volume within a radius of 0.81 Mpc (the projected virial radius from Equation~\ref{eq:virial_radius}). To estimate uncertainties, we perform a Monte Carlo sampling of the normalization parameter obtained from spectral fitting. Since the normalization must be non-negative, we draw samples from a truncated normal distribution \citep{2020NatMe..17..261V}, centered on the best-fit normalization with a standard deviation corresponding to its $1\sigma$ error.

We obtain a best-fit $n_e=6.35^{+2.33}_{-2.63}\times 10^{-5}$cm$^{-3}$ and $M_{\text{gas}} = 1.74^{+0.64}_{-0.72} \times 10^{12}$ M$_{\odot}$. As a result, the $1\sigma$ upper limit on the gas mass fraction is $f_{\text{gas}}\sim0.05$ for a weak-lensing mass of $M_{500}=0.6\times M_{200}=7.8^{+5.4}_{-5.4}\times 10^{13} \rm M_{\odot}$. In addition, we use the $L_{\rm X}-M_{\text{gas}}$ scaling relation from \citet{2015A&A...573A.118L}. Adopting a similar Monte Carlo approach by utilizing the best-fit X-ray luminosity in the $0.1-2.4$\,keV band ($=1.40^{+2.09}_{-1.40} \times 10^{42}$ erg s$^{-1}$), we obtain a $1\sigma$ upper limit of $M_{\text{gas}} = 1.4 \times 10^{12}$ M$_{\odot}$. The corresponding upper limit on gas mass fraction ($f_{\text{gas}}$) is $\sim0.02$. These estimates consistently suggest a low gas mass ($M_{\text{gas}}\lesssim1-3 \times 10^{12}\rm M_{\odot}$) and, consequently, a low gas mass fraction ($f_{\text{gas}}\lesssim0.02-0.05$) within the cluster, placing it at the lower end of the $f_{\text{gas}}$ distribution even for low X-ray surface brightness clusters \citep{2010MNRAS.407..263A}.

\begin{figure}[ht]
%\centering
\includegraphics[width=\columnwidth]{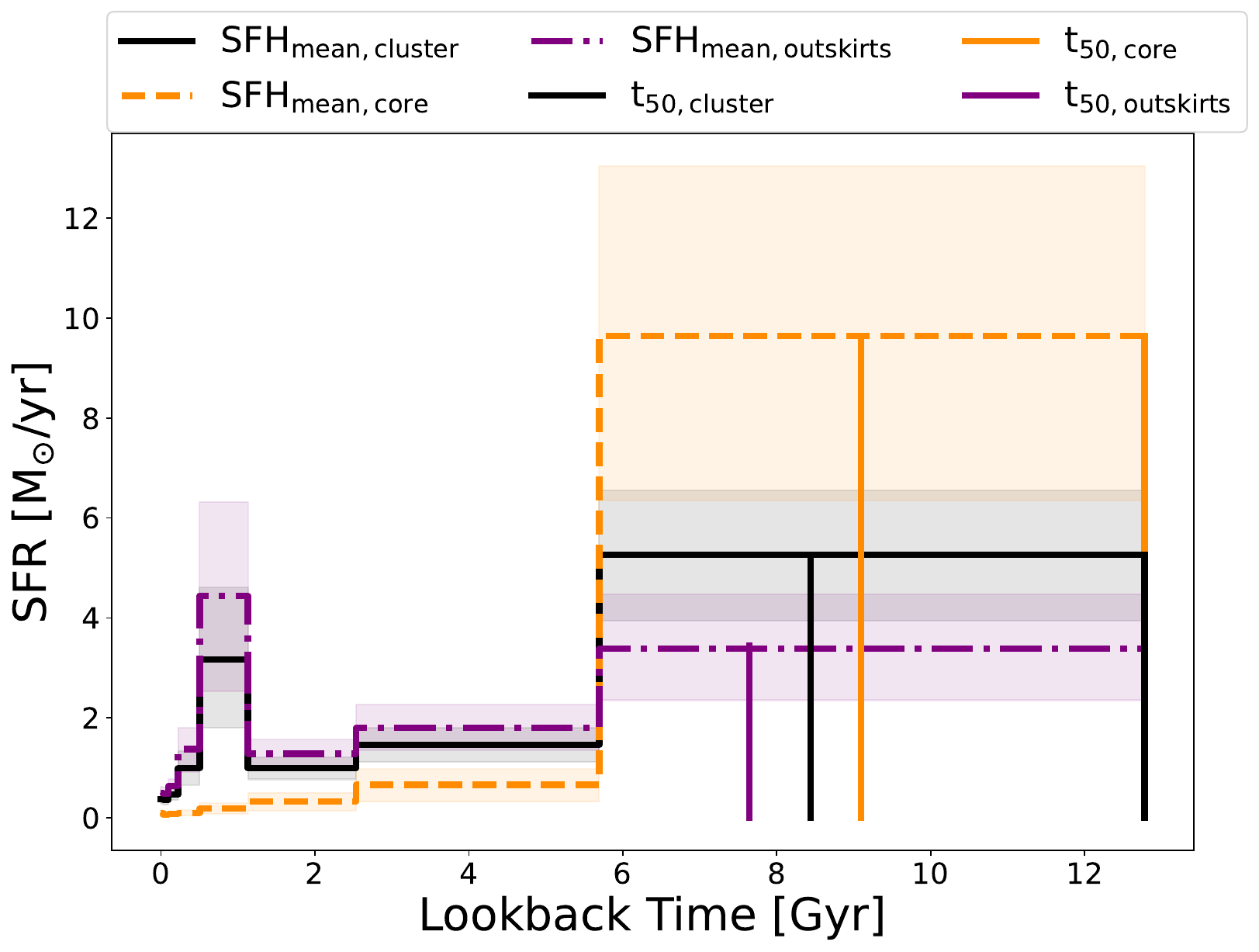}
\caption{SFH of the blue galaxy cluster, SDSS-C4\,3028. The solid black, dashed orange and dotted-dashed purple lines represent the mean SFHs of all member galaxies, of galaxies residing in the cluster core, and the outskirts, respectively. The vertical solid black, orange and purple lines represent the mass-weighted ages (t$_{50}$) inferred from the mean SFHs of the cluster, core and outskirts, respectively. Overall, the outskirts are more active in SF when compared to the core galaxies over last $\sim1$\,Gyr. Additionally, on average, the galaxies within the core assembled half of their stellar masses $\sim1.5$\,Gyr earlier than their outskirts, consistent with the inside-out formation and quenching pattern.
}
\label{fig:SFHs}
\end{figure}

\begin{figure}[h!]
\centering
\includegraphics[width=\linewidth]{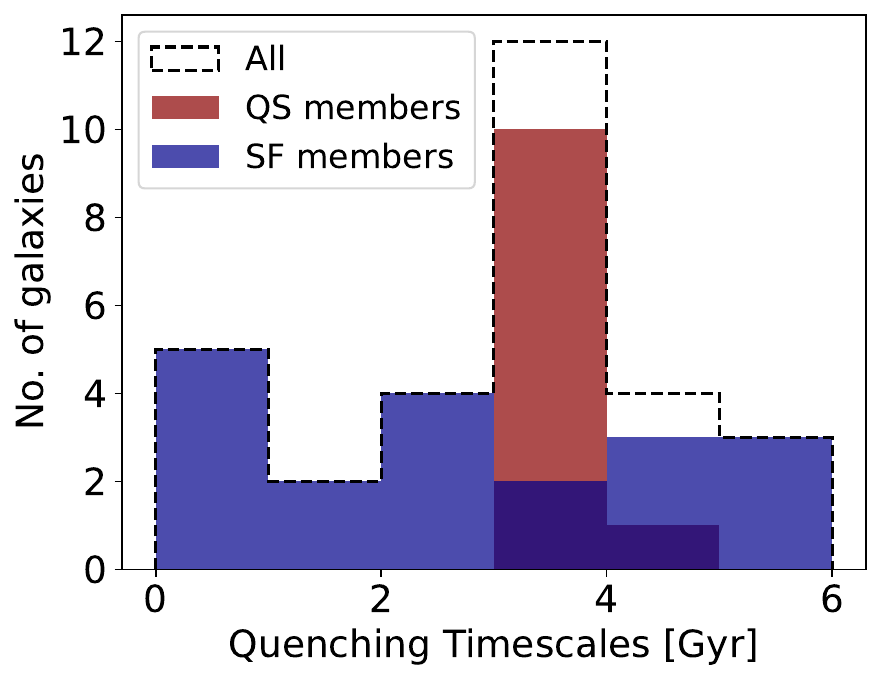}
\caption{Histogram of quenching timescales (the time over which a galaxy halves its SFR), derived from the SFHs of all (black dashed), SF (blue) and QS (red) member galaxies of SDSS-C4\,3028. All QS galaxies have a relatively long t$_{\rm quench}$ of $>3$\,Gyr, favoring slow quenching
mechanisms such as galaxy-galaxy interaction or strangulation.} 
\label{fig:Quenching timescales}
\end{figure}

\subsection{X-ray properties of the member galaxies} \label{sec:Members}
Our analysis of the X-ray image of SDSS-C4\,3028 indicates no diffuse emission on the cluster scale (Figure~\ref{fig:X-ray image}). However, there appears to be diffuse X-ray emission that spatially aligns with a few member galaxies. 
In Appendix Figure~\ref{fig:SB}, we present the surface brightness profiles of member galaxies with extended X-ray emission, extracted from the combined mosaic, background-subtracted and exposure-corrected \xmm image in the energy range of 0.5--2.0\,keV. The average local background level is represented by pink dashed lines in each of the plots. We identified 4 galaxies with diffuse emission, whose surface brightness is more extended than that of a nearby point source. 
To estimate the X-ray luminosities of individual galaxies, we use WebPIMMs\footnote[6]{\url{https://heasarc.gsfc.nasa.gov/cgi-bin/Tools/w3pimms/w3pimms.pl}} to convert count rates to flux (erg\,s$^{-1}$\,cm$^{-2}$) in the desired energy band. We assume a thermal {\tt apec} model with absorption fixed at $N_{\rm H}\,=\,2 \times 10^{20}$\,cm$^{-2}$, abundance $\rm Z\,=\,0.4\,Z_{\odot}$, and temperature $kT\,=\,0.5$\,keV \citep{2015ApJ...806..156S, 2013ApJ...766...61S}. The average local background level is estimated using a circular region with a radius of $\sim11.6^{\prime}$ (820.6\,kpc), centered at the aimpoint, excluding both the point sources and the member galaxies. Among the 4 galaxies with diffuse emission, ID~15 stands out for its elevated $L_\mathrm{X}(0.5-2.0)$~keV of $6.65^{+0.93}_{-0.93}\times 10^{40}\rm\,erg\,s^{-1}$ (see Table~\ref{tab:Properties}) with a best-fit $L_{\rm X}(0.1-2.0~{\rm keV})/L_{\rm K}$ ratio of $3.00\times 10^{30}\rm\,erg\,s^{-1}L_{\rm K_{\odot}}^{-1}$comparable to the dominant galaxies of nearby galaxy groups \citep{2015ApJ...806..156S}, while the $L_\mathrm{X}/L_\mathrm{K}$ ratios of others are consistent with cluster or group non-central member galaxies. We did not detect the corresponding X-ray emission in the 2.0--10.0\,keV band for ID~15, demonstrating this feature is soft. However, with the current \xmm observations, we cannot rule out the presence of a soft AGN in addition to the diffuse ISM.

We identify AGN by looking for any X-ray point source located within a maximum of 0.3$^{\prime}$ (21.2\,kpc) of the optical nucleus of any member galaxy. Three point-like X-ray sources are found to be associated with a member galaxy. Assuming a power-law model with the absorbing column fixed to $N_{\rm H} = 1.0 \times 10^{21}$ cm$^{-2}$ \citep{2016A&A...594A.116H} and a photon index of $\Gamma = 1.8$, we consider a source to be an AGN if its X-ray luminosity exceeds $10^{41}\rm\,erg\,s^{-1}$ in the 2.0--10.0\,keV band. This results in three X-ray AGN, corresponding to an occupation fraction of $\sim0.1$. Two of the AGN are associated with galaxies ID~14 and ID~30. The brightest AGN is spatially coincident with two SF galaxies (ID~1 and ID~4) that are located only 8.5\,kpc apart and are likely in the process of merging on the basis of their optical image. This point source is very luminous in hard X-rays, with $L_{\rm X}\text{(2-10 keV)}\sim2.09 \times 10^{42}\rm\,erg\,s^{-1}$, while its soft band X-ray luminosity is $L_{\rm X}\text{ (0.5-2 keV)}=8.45 \times 10^{40}$\,erg s$^{-1}$. Although the \xmm observation cannot resolve whether this brightest X-ray source is a single or a dual AGN, the detection suggests that it may have been triggered by an ongoing galaxy merger. 

\subsection{Star-formation histories of member galaxies}\label{Results:sfhs}

We present the SFH of cluster member galaxies individually and collectively in Figures~\ref{fig:Members} and \ref{fig:SFHs}. 
More than half (IDs~1, 4, 5, 6, 7, 8, 9, 11, 23, 25, 27, 29, 30) of the galaxies residing in the outskirts (R~$>300$\,kpc) display an increase in SF activity within the last 1--2\,Gyr, possibly governed by newly accreted blue satellites.
We derive a number of time scales to gain insight into the physical processes driving galaxy evolution.
We calculate the mass-weighted age, $\rm t_{50}$, defined as the lookback age when a galaxy has assembled half of its stellar mass. We find galaxies residing in the core (R$<300$\,kpc) assembled half of their masses $\sim1.48^{+0.87}_{-0.58}$\,Gyr before the galaxies on the outskirts (see Figure~\ref{fig:SFHs}). The SF timescales ($\rm t_{\rm SF}$), defined for each galaxy as length of time between the formation of 10\% and 90\% of its stellar mass, differ between the core and outskirts populations. Core galaxies formed the bulk of their stellar mass earlier, $\sim\,4.37^{+1.08}_{-0.50}$~Gyr before the outskirts, with $\rm t_{\rm SF, core}=5.89^{+0.24}_{-0.13}$\,Gyr and $\rm t_{\rm SF, outskirts}=10.48^{+0.26}_{-1.18}$\,Gyr (see Figure~\ref{fig:SFHs}). The ``quenching timescale'' ($\rm t_{\mathrm{quench}}$) is defined as the time it takes for a galaxy to reduce its SFR by half. We call it ``sudden quenching'' if the quenching time is $\lesssim1$\,Gyr. All QS galaxies have a relatively long $\rm t_{\mathrm{quench}}$ of $>3$\,Gyr (see Figure~\ref{fig:Quenching timescales}).
%^{+1800}_{-660}
\begin{figure*}
\centering
\includegraphics[width=\linewidth]{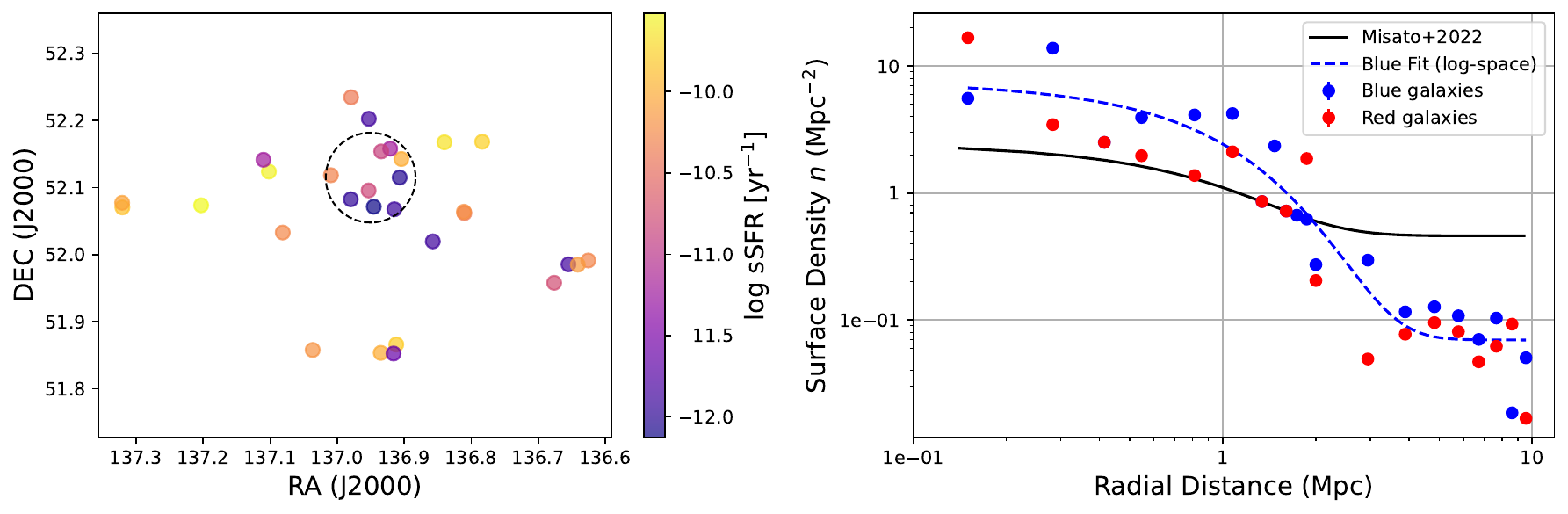}
\caption{\textit{Left:} Distribution of member galaxies, color-coded by sSFR. The galaxies tend to exhibit negligible sSFR towards the core (marked by a dashed black circle) and higher sSFR towards the outskirts. \textit{Right:} The surface density profiles of the blue (blue scatter points) and red (red scatter points) galaxies in and around SDSS-C4\,3028. Both axes are shown in logarithmic scale. The dashed blue curve  shows the best-fit surface density profile for blue galaxies in SDSS-C4\,3028. The fit was performed in logarithmic space.  The fit uses the same functional form as \citealt{Misato_2022}, $n(r) = n_0[1 -\rm tanh (r/\sigma)] + constant$, with best-fit parameters, $n_0 = 7.64\pm2.34~\rm Mpc^{-2}$, $\sigma =1.18\pm0.17~\rm Mpc$ and $\rm constant = 0.07\pm0.02~\rm Mpc^{-2}$. For comparison, the solid black curve represents the best-fit model for emitter galaxies in 43 blue clusters at $z=0.84$ from \citealt{Misato_2022}. SDSS-C4\,3028 has a more concentrated galaxy distribution than high-redshift blue clusters.}
\label{fig:sSFR_var}
\end{figure*}

\section{Discussion} \label{sec:Discussion}

\subsection{Where does the Gas in the Cluster Go?} \label{Analogy}

SDSS-C4\,3028 shows no significant diffuse X-ray emission, strongly suggesting the absence of an ICM, atypical for a cluster of this mass (Section~\ref{sec:mass} and \ref{sec:ab_ICM}). Moreover, the cluster exhibits an extraordinarily high fraction ($\sim0.63 \pm 0.08$) of SF galaxies among the total member galaxies. The fraction of SF galaxies stands in stark contrast to observations of nearby galaxy clusters, $>\,4\sigma$ higher than the median fraction of $\sim0.2$ for clusters at similar redshifts and virial masses \citep{Hashimoto_2019}. 

At first glance, the lack of an ICM and the elevated SF fraction makes SDSS-C4\,3028 resemble high redshift blue galaxy clusters, such as those at $z=0.84$ studied by \citet{Misato_2022}. The authors suggested that the hot gas in these young blue clusters has not yet been compressed and heated sufficiently to emit in X-rays, in line with the notion that X-ray underluminous clusters might still be in the process of formation \citep{2006MNRAS.373..653R, 2007A&A...461..397P}. To test this hypothesis for SDSS-C4\,3028, we compare its average number density with that of the blue clusters from \citealt{Misato_2022} (see Figure~\ref{fig:sSFR_var}-right), which reveals that SDSS-C4\,3028 has more than twice the number density observed in high redshift blue clusters. SDSS-C4\,3028 is therefore unlikely to be in an early evolutionary stage.

We also consider the possibility that a greater fraction of the primordial gas may have been converted into stars instead of remaining in the ICM. The cluster’s mass-to-light ratio would be significantly lower than that of typical massive clusters, reflecting an unusually high stellar baryon fraction. We estimate a stellar baryon fraction ($\sim0.01$) similar to what is found in local clusters of similar total mass \citep{2010MNRAS.407..263A}, suggesting that this scenario is unlikely.

Another alternative scenario would be that the cluster is unable to retain its ICM. One key mechanism that could expel hot gas from a cluster is the non-gravitational heating caused by a strong feedback mechanism, such as by an AGN \citep{2022A&A...666A..22R}. For instance, in a study by \citet{2022A&A...666A..22R}, the authors analyzed $\sim1800$ galaxy clusters from a high-resolution \textit{Magneticum} hydrodynamic cosmological simulation \citep{2016MNRAS.463.1797D} and found that older clusters tend to be gas poor due to AGN feedback processes, which eject gas from the halo over time. Although the impact of feedback provides a suggestive and physically grounded scenario for the existence of an underluminous cluster, the high fraction of blue galaxies in this cluster remains puzzling. %, particularly given that other nearby X-ray faint clusters are not necessarily dominated by SF galaxies.

Another noteworthy observation is the spatial distributions of SF and QS galaxies (see Section~\ref{Structural Analysis} and \ref{Results:sfhs}). $54\pm9\%$ of the red, QS galaxies and $16\pm5\%$ of the blue, SF galaxies, are located inside the core with a radius of $\sim300$\,kpc. Therefore, the majority of the red galaxies are in the core (redder core with a $f_{\rm SFG}$ of $\sim0.33$), while the majority of the blue ones are in the outskirts (bluer outskirts with a $f_{\rm SFG}$ of $\sim0.76$). Furthermore, the bluer outskirts correlate with an increasing specific SFR (sSFR) with radial distance (see the left plot of Figure~\ref{fig:sSFR_var}). The median sSFR of the core population (log sSFR$_{\rm median, core}$/yr = $-11.2$) is an order of magnitude smaller than that for the outskirts population ($\log \rm sSFR_{\rm median, outskirts}$/yr = $-10.2$). The dense core of red galaxies and the high density of galaxies within the cluster (see Figure~\ref{fig:sSFR_var}) suggest that the system underwent an early phase of collapse, which would have led to the formation of a more massive cluster had the collapse continued uninterrupted. However, if the infall paused for an extended period, AGN activity triggered by the initial collapse may have efficiently cleared out a substantial fraction of the gas--at least from the relatively low-mass core at that stage. Subsequently, the blue, SF galaxies may have formed in lower-mass halos and fallen into the system after most of the gas had already been expelled \citep{2020RAA....20..207W}. This is in line with the possible LOS merger in SDSS-C4\,3028 (Section~\ref{Structural Analysis}) and raises the possibility that the cluster mass could be over-estimated due to projection effects. If the system consists of two or more lower-mass groups aligned along our LOS, both the weak-lensing and dynamical mass estimates would be biased high. Although the velocity distribution does not show significant deviations from a single Gaussian, improved sampling of galaxy velocities with more complete spectroscopic coverage in the future could reveal potential substructures.

In summary, the key elements shaping this system’s evolution appear to be an early, rapid collapse, a subsequent pause in halo mass growth, and a later phase of infall and merger of sub-clusters. In the context of the spherical collapse model, achieving such a collapse history would require an overdense core surrounded by an underdense region, slowing down the next stage of collapse, followed by an additional overdense region that triggered the more recent phase of infall. While the spherical model is a simplification, it illustrates a working scenario that is consistent with all the findings. The formation of SDSS-C4\,3028 may require a complex and atypical collapse history, which may explain why clusters like SDSS-C4\,3028 are not commonly observed.

\subsection{The role of ram pressure stripping} \label{RPS}
 
Among the various processes responsible for galaxy quenching, RPS is emphasized as a key mechanism in nearby rich galaxy clusters. This phenomenon occurs when galaxies navigate the cluster's gravitational potential with high velocities and their ISM experience a resistive force as they traverse the ICM \citep{gunn1972infall, Boselli_2022}. The strength of ram pressure is $P_{\mathrm{ram}}$ $\sim\rho_{\mathrm{ICM}}v^2$, where $\rho_{\mathrm{ICM}}$ is the ICM density and $v$ is the relative velocity between the galaxy and the ambient ICM. Once the ram pressure exceeds the gravitational restoring force per unit area, the stripping of galactic gas may begin. Since the gravitational restoring force varies across the galactic disk, the outer, less tightly bound regions, are stripped first, while the inner disk gas typically requires higher ram pressure to be removed. This process may enrich the ICM over time \citep{Schindler_2008}. 

Since ram pressure is directly proportional to the density of the ICM, the absence of a sufficiently dense gas in SDSS-C4\,3028 (Section~\ref{sec:ab_ICM} and \ref{sec:Members}) may imply low ram pressure, and therefore ineffective RPS. Furthermore, the long observed quenching times (t$_{\rm quench}>3$\,Gyr; Section~\ref{Results:sfhs}) for quiescent galaxies in this cluster imply slow quenching processes such as galaxy-galaxy interaction or strangulation, while the quenching time expected for RPS is $\lesssim1$~Gyr \citep[see][for a detailed review]{Boselli_2022}. Both findings imply that quenching is not due to RPS in this cluster. The absence of RPS may be responsible for the observed elevated fraction of SF galaxies, highlighting the probable role of RPS in quenching member galaxies in more typical clusters. 

\section{Conclusions} \label{sec:Conclusions}

In this paper, we present our analysis of a new $\sim\,20$\,$\mathrm{ ksec}$ \xmm observation of SDSS-C4\,3028, a galaxy cluster with a dark matter halo mass of $>~10^{14}\rm M_{\odot}$ in the local Universe ($z = 0.061$), notable for its unusually high fraction of SF galaxies \citep{Hashimoto_2019}. We complement our \xmm analysis with a WL analysis using Subaru/HSC observations as well as SED fitting to construct SFHs for the cluster member galaxies. Our main findings and conclusions are as follows:

\begin{itemize}

    \item The cluster is comprised of 30 spectroscopically-confirmed member galaxies with $r$-band absolute magnitudes of $M_r \lesssim -20.1$. With 19 SF and 11 QS galaxies, we derive a SF fraction of $f_{\text{SFG}} = 0.63 \pm 0.08$, higher than previously reported. 
    
    \item The uniform diffuse emission is fully consistent with the astrophysical X-ray background, with no X-ray brightness enhancement associated with the cluster. The stringent $3\sigma$ upper limit of $L_{\rm X}\text{ (0.1-2.4 keV)}\,\sim\,7.7\,\times\,10^{42}$\,erg\,s$^{-1}$ confirms the lack of an ICM, suggesting a low intracluster gas density. AGN feedback may have expelled much of the ICM early in the cluster’s formation.
    
     \item The lack of an ICM implies the absence of RPS at play, which is likely the reason for the elevated fraction of SF galaxies in SDSS-C4\,3028. Conversely, this highlights the role of RPS in quenching galaxies in more typical clusters with substantial ICM. 

     \item  More than half of the QS galaxies lie within the central $\sim300$~kpc. We propose that the quenching of these galaxies is a result of mechanisms like galaxy-galaxy interactions or strangulation, as well as stellar and AGN feedback. The SFHs show that the galaxies in the outskirts are more actively SF over last 1\,Gyr than the core galaxy population, while the redder core assembled half of its stellar mass $\sim1.5$\,Gyr earlier than the bluer outskirts. 
     The presence of a redder core and bluer outskirts suggests an ``inside-out quenching''.      
\end{itemize}
SDSS-C4\,3028 provides valuable constraints on theories and cosmological simulations in which the baryonic content is shaped by stellar and AGN feedback during cluster evolution and large-scale structure formation.
It also offers a rare opportunity to study the impact of mechanisms of quenching cluster galaxies other than RPS. 
SDSS-C4\,3028 may represent a broader population of blue clusters in the nearby Universe, which have rarely been discovered, probably because most galaxy cluster surveys trace the ICM content (through X-ray or Sunyaev–Zeldovich observations) or the red sequence member galaxies \citep[e.g., RedMaPPer;][]{2014ApJ...783...80R}. To enhance the discoverability of such blue clusters, future optical surveys may incorporate methods beyond these traditional selection techniques, such as spectroscopic clustering or tracers of SF galaxies. 

\section{acknowledgments}
\twocolumngrid
We thank the anonymous reviewer for their constructive feedback, which helped improve the manuscript. S.J. and Y.S. acknowledge the support of NASA grant 80NSSC22K0856. This work uses observations obtained with \xmm, an ESA science mission with instruments and contributions directly funded by ESA Member States and NASA. H. Cho acknowledges support for the current research from the National Research Foundation of Korea (NRF) under the programs RS-2022-NR070872 and RS-2023-00219959. This work was supported in part by the K-GMT Science Program (PID: GEMINI-KR-2022B-011) of Korea Astronomy and Space Science Institute (KASI). This research is based in part on data collected at the Subaru Telescope (PID: S22B-TE011-GQ), via the time exchange program between Subaru and the international Gemini Observatory, a program of NSF's NOIRLab. The Subaru Telescope is operated by the National Astronomical Observatory of Japan. We are honored and grateful for the opportunity of observing the Universe from Maunakea, which has the cultural, historical, and natural significance in Hawaii. This research used data obtained with the Dark Energy Spectroscopic Instrument (DESI). DESI construction and operations is managed by the Lawrence Berkeley National Laboratory. This material is based upon work supported by the U.S. Department of Energy, Office of Science, Office of High-Energy Physics, under Contract No. DE–AC02–05CH11231, and by the National Energy Research Scientific Computing Center, a DOE Office of Science User Facility under the same contract. Additional support for DESI was provided by the U.S. National Science Foundation (NSF), Division of Astronomical Sciences under Contract No. AST-0950945 to the NSF's National Optical-Infrared Astronomy Research Laboratory; the Science and Technology Facilities Council of the United Kingdom; the Gordon and Betty Moore Foundation; the Heising-Simons Foundation; the French Alternative Energies and Atomic Energy Commission (CEA); the National Council of Humanities, Science and Technology of Mexico (CONAHCYT); the Ministry of Science and Innovation of Spain (MICINN), and by the DESI Member Institutions: \url{www.desi.lbl.gov/collaborating-institutions}. The DESI collaboration is honored to be permitted to conduct scientific research on I'oligam Du'ag (Kitt Peak), a mountain with particular significance to the Tohono O'odham Nation. Any opinions, findings, and conclusions or recommendations expressed in this material are those of the author(s) and do not necessarily reflect the views of the U.S. National Science Foundation, the U.S. Department of Energy, or any of the listed funding agencies.

\appendix
\setcounter{figure}{0}
\setcounter{table}{0}
\renewcommand{\thefigure}{\Alph{section}\arabic{figure}}
\renewcommand{\thetable}{\Alph{section}\arabic{table}}
\section{Member galaxies}

\begin{deluxetable*}{lccccccccccc}
\tabletypesize{\scriptsize}
\tablewidth{0pt} 
\tablecaption{Properties of member galaxies \label{tab:Properties}}
\tablehead{
\colhead{ID} & \colhead{RA} & \colhead{DEC} & \colhead{$z_{\rm spec}$} & \colhead{$L_\mathrm{X}$} & \colhead{$L_\mathrm{X}$} & \colhead{$L_\mathrm{X}$} & %\colhead{log $L_\mathrm{K}$} &
\colhead{log SFR} & \colhead{log $M_{\star}$} & \colhead{SF/QS} & \colhead{Morphology}\\
\colhead{ } & \colhead{ } & \colhead{ } & \colhead{ } & \colhead{(0.5--2.0 keV)} & \colhead{(0.5--2.0 keV)} & \colhead{(2.0--10.0 keV)} & \colhead{ } & \colhead{ } & \colhead{ } %& \colhead{ }
& \colhead{ }\\
\colhead{ } & \colhead{ } & \colhead{ } & \colhead{ } & \colhead{Diffuse} & \colhead{Point sources} & \colhead{Point sources} & \colhead{ } & \colhead{ } & \colhead{ } & %\colhead{ } & 
\colhead{ }\\
\colhead{ } & \colhead{$(J2000.0)$} & \colhead{$(J2000.0)$} & \colhead{ } & \colhead{(10$^{40}$ erg s$^{-1}$)} &\colhead{(10$^{40}$ erg s$^{-1}$)} & \colhead{(10$^{41}$ erg s$^{-1}$)} &  \colhead{(yr$^{-1}$)} & \colhead{(M$_{\odot}$)} & %& \colhead{(erg s$^{-1}$)}
\colhead{ } &  \colhead{ }
}
\startdata 
 1 & 136.8102 & 52.0621 & 0.05955 & - & 8.45 $\pm$ 0.98 & 20.90 $\pm$ 1.09 &  0.19 $\pm$ 0.43 & 10.57 $\pm$ 0.10 & SF* & - \\  
 2 & 136.9530 & 52.0959 & 0.06049 & 1.84 $\pm$ 0.53 & - &  - &   0.15 $\pm$ 0.47 & 10.97 $\pm$ 0.10 & SF & Spiral \\  
 3 & 136.9042 & 52.1427 & 0.06040 & - & - & - & 0.40 $\pm$ 0.20 & 10.40 $\pm$ 0.09 & SF & Spiral  \\  
 4 & 136.8109 & 52.0641 & 0.06024 & - & 8.45 $\pm$ 0.98 &  20.90 $\pm$ 1.09 &  0.14 $\pm$ 0.50 & 10.34 $\pm$ 0.10 & SF* & - \\  
 5 & 136.8400 & 52.1676  & 0.06102 & - & - & - & 0.20 $\pm$ 0.28 & 9.86 $\pm$ 0.07 & SF & Spiral\\ 
 6 & 136.9118 & 51.8662 & 0.06297 & - & - & - & 0.64 $\pm$ 0.15 & 10.44 $\pm$ 0.09 & SF & Spiral\\  
 7 & 136.9349 & 51.8536 & 0.06170 & - & - &  - & 0.18 $\pm$ 0.16 & 10.18 $\pm$ 0.10 & SF & - \\ 
 8 & 137.0811 & 52.0331 & 0.05971 & - & - & - & -0.40 $\pm$ 0.47 & 9.86 $\pm$ 0.10 & SF &  - \\  
 9 & 136.7836 & 52.1685 & 0.06232 & - & - & - & 0.62 $\pm$ 0.27 & 10.39 $\pm$ 0.08 & SF & Spiral \\ 
 10 & 137.0092 & 52.1186 & 0.06276 & 1.54 $\pm$ 0.47  
 & -  & - & 0.20 $\pm$ 0.19 & 10.51 $\pm$ 0.09 & SF & Spiral \\  
 11 & 136.6254 & 51.9914 & 0.06340 & - & - & - & 0.41 $\pm$ 0.32 & 10.69 $\pm$ 0.10 & SF & Spiral\\  
 12 & 136.6763 & 51.9581 & 0.06274 & - & - & - & -0.04 $\pm$ 0.53 & 10.73 $\pm$ 0.11 & SF & Spiral \\  
 13 & 136.9159 & 51.8527 & 0.06253 & - & - & - & -0.86 $\pm$ 0.67& 10.77 $\pm$ 0.08 & QS & Elliptical \\  
 14 & 136.6549 & 51.9857 & 0.06158 & - & 1.38 $\pm$ 0.53 & 0.91 $\pm$ 0.41 & -0.96 $\pm$ 0.75 & 10.91 $\pm$ 0.09 & QS* & - \\  
 15 & 136.9067 & 52.1152 & 0.05985 & 6.65 $\pm$ 0.93 & - & -  
 & -0.92 $\pm$ 0.76& 11.11 $\pm$ 0.09 & QS &  Elliptical \\
 16 & 136.9147 & 52.0678 & 0.05755 & - & - & -  
 & -1.40 $\pm$ 0.70 & 10.51 $\pm$ 0.09 & QS & Elliptical \\
 17 & 136.9796 & 52.0827 & 0.06255 & - & - & -  
 & -1.19 $\pm$ 0.68 & 10.78 $\pm$ 0.08 & QS & - \\
 18 & 136.9453 & 52.0716 & 0.06234 & - & - & -  
 & -0.76 $\pm$ 0.73 & 11.36 $\pm$ 0.08 & QS & Spiral \\
 19 & 136.8573 & 52.0199 & 0.06021 & - & - & -  
 & -1.36 $\pm$ 0.69 & 10.52 $\pm$ 0.09 & QS & -  \\
 20 & 136.9209 & 52.1582 & 0.05734 & - & - & -  
 & -1.08 $\pm$ 0.65 & 10.13 $\pm$ 0.11 & QS &  - \\
 21 & 136.9342 & 52.1540 & 0.06115 & - & - & -  
 & -0.58 $\pm$ 0.46 & 10.30 $\pm$ 0.10 & QS & Spiral\\
 22 & 137.0365 & 51.8580 & 0.06419 & - & - & -  
 & 0.29 $\pm$ 0.37 & 10.49 $\pm$ 0.10 & SF & - \\
 23 & 137.2028 & 52.0737 & 0.06290 & - & - & -  
 & 0.81 $\pm$ 0.09 & 10.33 $\pm$ 0.10 & SF & - \\
 24 & 136.9527 & 52.2027 & 0.06086 & 1.00 $\pm$ 0.59 & - & -  
 & -0.69 $\pm$ 0.70 & 11.18 $\pm$ 0.09 & QS & - \\
 25 & 137.3203 & 52.0706 & 0.06145 & - & - & -  
 & 0.16 $\pm$ 0.24 & 10.02 $\pm$ 0.08 & SF & - \\
 26 & 137.3207 & 52.0774 & 0.06184 & - & - & -  
 & 0.37 $\pm$ 0.30 & 10.45 $\pm$ 0.09 & SF & Spiral \\
 27 & 137.1018 & 52.1239 & 0.06105 & - & - & - 
 & 0.14 $\pm$ 0.25 & 9.71 $\pm$ 0.07 & SF & Spiral \\
 28 & 137.1099 & 52.1416 & 0.06132 & - & - & - 
 & -1.14 $\pm$ 0.60 & 10.08 $\pm$ 0.09 & QS & - \\
 29 & 136.6410 & 51.9852 & 0.06412 & - & - & - 
 & -0.05 $\pm$ 0.26 & 10.08 $\pm$ 0.09 & SF & Spiral \\
 30 & 136.9791 & 52.2348 & 0.06117 & - & 5.37 $\pm$ 0.93 & 1.59 $\pm$ 0.55 & 0.22 $\pm$ 0.38 & 10.61 $\pm$ 0.09 & SF* & Spiral \\
\enddata
\tablecomments{
    Please refer to Figure~\ref{fig:Members} for the optical counterparts with the same IDs}
\tablenotetext{*}{X-ray AGN (see Section~\ref{sec:Members})}
\end{deluxetable*}
\section{Surface brightness profiles}
\begin{figure*}[h!]
    \includegraphics[width=1.0\textwidth]{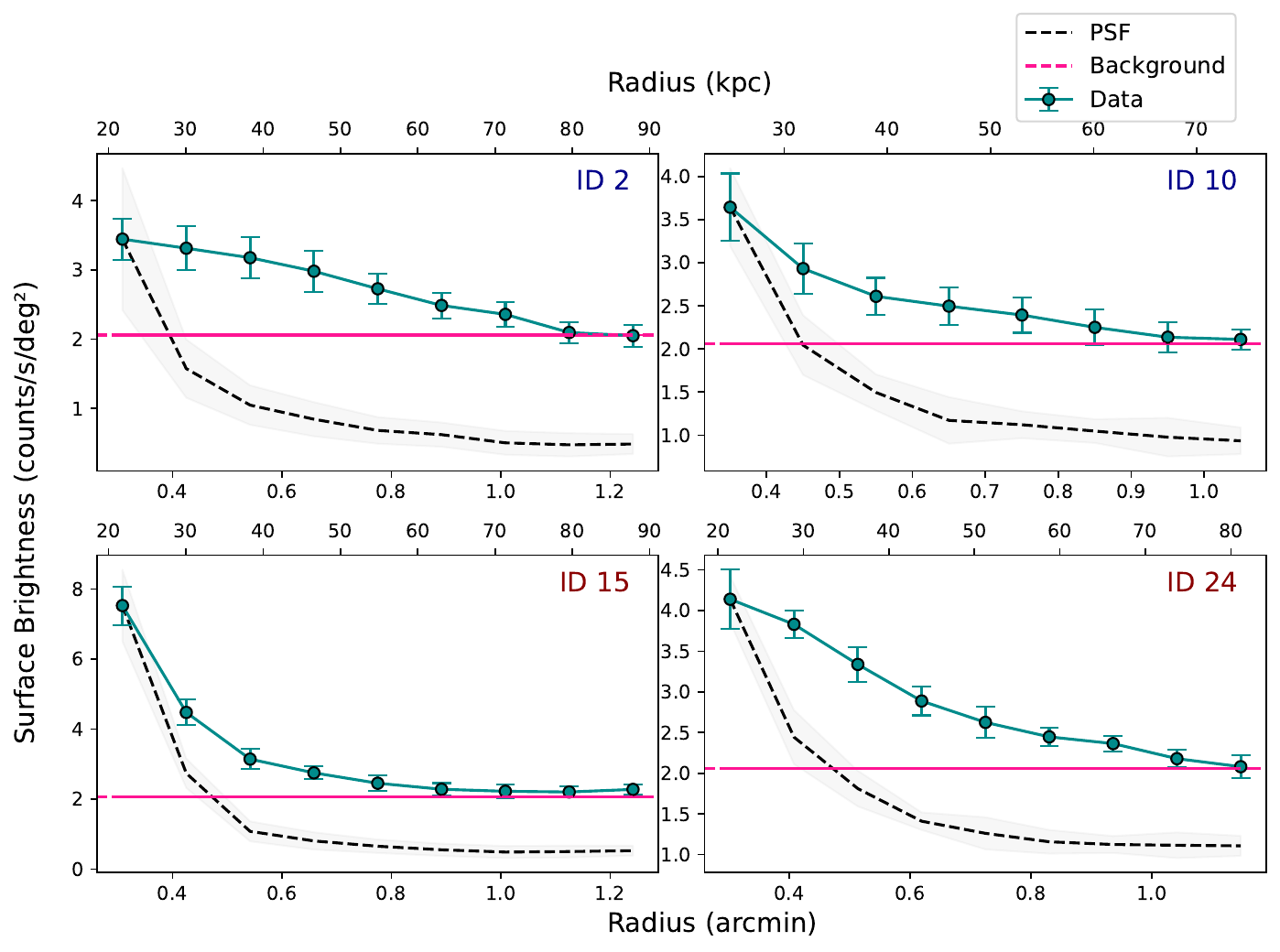}
    \caption{X-ray surface brightness profiles, centered on four cluster members with diffuse X-ray emission detected. The numbers in blue and red represent the IDs assigned to the star-forming and quiescent galaxies in this work, respectively. The horizontal pink line marks the average local background level. The PSF (dashed black line) denotes the point spread function profile of a nearby point source.
    \label{fig:SB}}
\end{figure*}

Figure~\ref{fig:SB} presents the surface brightness profiles of cluster members with detected diffuse X-ray emission. To ensure reliable measurements and minimize PSF broadening, we show profiles only for galaxies within a 10$^{\prime}$ radius around the \xmm\ aimpoint.

\bibliography{main}{}
\bibliographystyle{mnras_custom}
\end{document}